\PassOptionsToPackage{table}{xcolor}
\documentclass[]{fairmeta}

\title{Could LLM Watermark Detection be Public?}

\author[1,2,*]{Georgios Milis}
\author[1]{Tom Sander}
\author[1]{Tom\'{a}\v{s} Sou\v{c}ek}
\author[2]{Heng Huang}
\author[1]{Pierre Fernandez}

\affiliation[1]{FAIR, Meta Superintelligence Labs}
\affiliation[2]{University of Maryland, College Park}

\contribution[*]{Work done during an internship at Meta}

\abstract{

Watermarking large language models is popular for tracing chatbot and agentic outputs, yet detectors remain unreleased since exposing them could let attackers do targeted edits with the detector's feedback. 
However, watermarks are already vulnerable to uninformed tampering attacks. 
We thus first quantify whether a public detector would be an additional liability in a deployment setting at varying levels of access, from token-level scores to a binary verdict. 
Second, we introduce a split-key public-private watermarking method that exposes one key through a public detector while keeping the other for full verification and forensics. 
An informed attacker can only move the public signal, creating an imbalance between public and private scores. 
We introduce a statistical test for this imbalance, and combine it with the full key verdict in a two-stage mechanism. 
Third, we evaluate the split-key method on a wide range of removal and forgery attacks, comparing the uninformed to detector-informed settings. 
Public detection improves removal only at small edit budgets, since plain rephrasing already strips the watermark at a lower quality cost, but it does enable forgery, which the private pipeline can identify. 
Overall, releasing half of the watermark enables transparency and interoperability, and tampering with the released half stays detectable. 
This bounds the provider's liability and questions the need to keep detectors fully private.

}

\correspondence{\email{milis@umd.edu}, \email{pfz@meta.com}}
\metadata[Code]{\url{https://github.com/facebookresearch/textseal}}

\usepackage[utf8]{inputenc}
\usepackage[T1]{fontenc}
\usepackage{url}
\usepackage{amsmath, amssymb}
\usepackage{bbm}
\usepackage{algorithm, algpseudocode}
\usepackage{booktabs}
\usepackage{graphicx}
\usepackage{wrapfig}
\usepackage{needspace}
\usepackage{placeins}   %
\usepackage{caption}
\usepackage{subcaption}
\usepackage{xcolor}
\usepackage{pifont}
\usepackage{xspace}
\usepackage{enumitem}
\usepackage[normalem]{ulem}
\usepackage{multicol, multirow, makecell}
\usepackage{array}
\newcolumntype{C}[1]{>{\centering\arraybackslash}p{#1}}

\definecolor{heatblue}{HTML}{0072B2}
\colorlet{heathue}{heatblue!25!white}
\newcommand{\hc}[1]{\cellcolor{heathue!#1}#1\%}
\colorlet{rowhl}{heatblue!12}

\newcommand{\eg}{{e.g}.\@ }
\newcommand{\ie}{{i.e}.\@ }

\renewcommand{\paragraph}[1]{\par\noindent\textbf{#1}}
\definecolor{tiernobox}{HTML}{E15759}
\definecolor{tierwhitebox}{HTML}{4E79A7}
\definecolor{tierblackbox}{HTML}{4FA8A1}
\newcommand{\cnobox}[1]{\textcolor{tiernobox}{\textbf{#1}}}
\newcommand{\cwhitebox}[1]{\textcolor{tierwhitebox}{\textbf{#1}}}
\newcommand{\cblackbox}[1]{\textcolor{tierblackbox}{\textbf{#1}}}
\providecommand{\textcite}[1]{\citet{#1}}

\begin{document}
\maketitle

\section{Introduction}

\begin{figure}[b!]
\centering
\vspace{0.3em}
\includegraphics[width=\linewidth, clip, trim={0 7.5in 3in 0}]{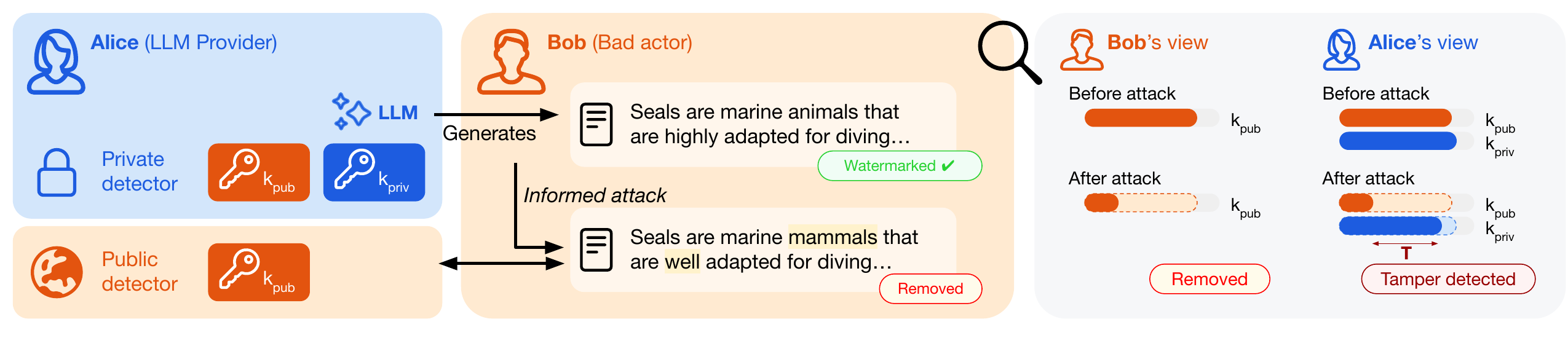}
\caption{\textbf{Overview of our approach.}
Alice, an LLM provider, splits the watermark signal across two keys, releasing a detector that uses only the public half $k_{pub}$ while keeping $k_{priv}$ private.
Bob may try to use the public detector for informed attacks that remove (or forge) the watermark more effectively than uninformed attacks.
Alice, who sees both keys, flags it as tampered from the imbalance $\mathbf{T}$ between the two scores.
Therefore, Bob has a better chance of evasion by using uninformed attacks, an already existing risk for Alice.
}
\label{fig:threat_model}
\end{figure}

Regulations increasingly require AI-generated content to be watermarked~\citep{EuropeanAIAct,EUCodeOfPractice2026,CaliforniaAITransparencyAct,IndiaWatermarking2025}, yet detection remains closed and fragmented, as most providers keep their detectors secret.
Verifying a text in the wild therefore depends on each provider granting access to its detector.
Cross-industry efforts like C2PA for multimedia do not apply to plain text.
Instead, if a provider released part of its detector publicly, any other platform could integrate it into its own detection, enabling interoperability.

However, sharing a detector is considered risky because it eases two attacks: \textit{removal} (scrubbing), which strips the watermark to hide LLM use and evade filters in schools or against spam and misinformation, and \textit{forgery} (spoofing), which adds watermarking signal to falsely blame a provider, for instance by inserting a name into a defamatory sentence or fabricating compliance evidence.
Importantly, both attacks are already possible without a detector, at a certain distortion.
Removal either rewrites the text globally or edits it locally, and often degrades quality, while forgery either only ``piggybacks'' on already-watermarked text~\citep{pang2024no} or reverse-engineers the watermark from a large corpus of watermarked outputs~\citep{jovanovic2024watermark}.
Yet personal agents increasingly communicate on their users' behalf, and a public detector would let recipients verify such agent-generated text.
On the other hand, a detector leaks information about the watermark and lets an attacker steer edits at lower distortion.
Watermarking research takes this for granted, and it is the standard reason detectors stay private, yet how much an attacker actually gains has not been measured.

Cryptographic approaches instead embed an unforgeable, publicly-verifiable signature.
The methods of \citet{fairoze2025publicly, pmlr-v258-fairoze25a} are either weakly robust, since detection needs a contiguous span of the text to survive, or intractable to construct for low-entropy text.
Since no cryptographic construction is yet robust, fast, and publicly verifiable at once, two questions remain open:
(a)~How much does public detection actually help an attacker at each level of access, and is the added liability large enough to justify keeping detectors secret?
(b)~Since cryptographic public detection is impractical, is there a feasible middle ground for statistical LLM watermarks?

To answer~(a), we compare an \textit{uninformed} attacker with an \textit{informed} one that queries the detector. We build informed removal and forgery attacks: we consider both rephrasing, via autoregressive regeneration, and in-place word edits steered by the detector.
We evaluate them at each level of access, from a fully transparent detector to the binary verdict a public API would return.
To answer~(b), we present an approach with no unforgeability claim but such that the strongest informed attacks remain \textit{detectable}: we split the watermark across independent keys and release only one for public detection, keeping the full set private.
This follows an old idea from asymmetric video watermarking, where publishing only part of the key keeps the watermark publicly detectable yet robust~\citep{hartung1997fast}, but its security is secrecy, not cryptography.
This key split exists in TextSeal~\citep{textseal2026}, which already routes each token to one of two keys, and extends to other schemes, like Maryland~\citep{kirchenbauer2023watermark} or SynthID-Text~\citep{dathathri2024scalable}.

Our main contributions are:
\textit{(i)} we are the first to measure what a public detector actually costs the provider that releases it, with attackers that query it at every level of access: scores or $p$-values enable strong informed attacks, while a binary verdict can help removal by indicating when to stop.
\textit{(ii)} We introduce a split-key construction with a two-stage detector, and a novel, calibrated tampering test that identifies the public-private imbalance an informed attacker creates.
\textit{(iii)} We show the split limits the threat caused by informed attacks, as our strongest informed forgery succeeds on almost all carrier texts yet the majority is caught, while informed removal helps only at small edit budgets.
In short, we establish that public detection carries a real but bounded liability, and enable future research on watermark interoperability and transparency.

\section{Background and Threat Model}

\subsection{LLM Watermarking}~\label{subsec:related_llmwm}

LLM watermarking alters the tokens' sampling in a key-dependent way to leave a statistical trace only the key holder can test for.
The \emph{key} is a fixed secret shared by sampler and detector: a pseudo-random function $\mathrm{PRF}$ maps it along with the $h$ preceding tokens to a score $r_t(v)$ per candidate token $v$, which the sampler combines with the LLM's next-token distribution $p_t$ to pick the next token $x_t$.

Schemes differ in how $r_t$ alters sampling, and divide into \emph{distortionary} and \emph{non-distortionary} families depending on whether that alteration changes the model's output distribution.
Distortionary methods bias sampling, \eg Maryland randomly partitions the vocabulary into green and red lists from the previous token and the secret key and adds $\delta$ to green logits \citep{kirchenbauer2023watermark}. 
Non-distortionary methods preserve the original model distribution in expectation over all possible keys. %
The exact definitions are in App.~\ref{app:distortion}.
In the rest, we mainly consider non-distortionary methods because they are the ones deployed in production systems~\citep{dathathri2024scalable,textseal2026,anthropic2026marking,li2026textgrain}.

Gumbel-max is a common single-token non-distortionary method \citep{aaronson2022}.
It is based on the Gumbel-softmax trick, and uses the pseudo-random $r_t(v)=\mathrm{PRF}(v; k,x_{t-h:t-1})\sim U(0,1)$ to choose the next token as $x_t=\arg\max_v r_t(v)^{1/p_t(v)}$.
This samples according to $p_t$ on expectation, but with key-dependent randomness. %
The detector recomputes $r_t(x_t)$ and tests the statistic $\mathbf{S}=\sum_t -\log(1-r_t(x_t))$, which follows $\Gamma(n,1)$ under the null hypothesis $H_0$ that the text is unwatermarked, and takes larger values under the alternative $H_1$.
While Gumbel-max samples deterministically, TextSeal restores diversity by instantiating it with 2 independent keys $k_1, k_2$ \citep{textseal2026}, which directly provides our public/private split.
Each token is sampled using one of the two keys selected at random with probabilities $\alpha$ and $1-\alpha$.
Since routing is unknown at detection time, scores under both keys are merged into a \emph{fused score} $\mathbf{S}=\alpha \mathbf{S}_1+ (1-\alpha)\mathbf{S}_2$.
Its null is a weighted sum of exponentials and is calibrated via a moment-matched Gamma to get a $p$-value $p_{\mathrm{fus}}$.

\begin{figure}[b!]
\centering
\vspace{0.8em}
\includegraphics[width=0.99\linewidth, clip, trim={0 7.1in 2.5in 0}]{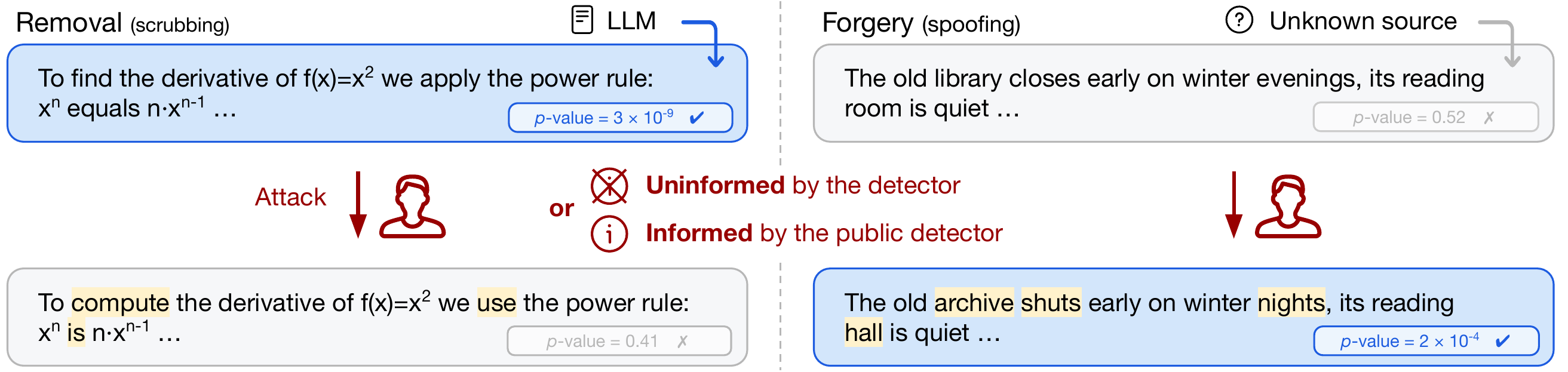}

\caption{
Removal and forgery attacks, under levels of detector access ranging from none to token-level scores.
}
\label{fig:scrub_spoof}
\end{figure}

\subsection{Threat Model}
\label{subsec:threat}

We assume an LLM provider that exposes a public watermark detector.
The attacker's goal is either to \textit{remove} the watermark so that AI-generated text evades detection, or to \textit{forge} it, \ie embed it into text the platform's model did not generate, as depicted in Fig.~\ref{fig:scrub_spoof}.
We also define: an \emph{uninformed} attacker, who has no detector access and tampers blindly, as commonly assumed in the watermarking literature, and an \emph{informed} attacker, who directly exploits the public detector to steer the attack.

We consider four scenarios of increasing detector transparency: the uninformed
\emph{no-box} (uninformed), where no public detector exists, as in current watermark deployments.
\emph{Black-box}, where the detector returns a binary watermark decision.
\emph{Gray-box}, where it returns the text's $p$-value,
and \emph{white-box}, a fully transparent detector that returns token-level scores.
The last three allow for \emph{informed} attacks of increasing strength.
\emph{White-box} is the most permissive setting so it upper-bounds the liability of publishing a detector, and we show that it is more advantageous for forgery.
\emph{Black-box} answers ``watermarked or not'', only serving as a stopping rule for attacks (Sec.~\ref{subsec:vulnerability}).
The gray-box tier behaves like the white-box one, since a token's score can be approximated from the $p$-value difference caused by its presence in the text (App.~\ref{app:less_transprent}).

\subsection{Attacks on a Public Detector}
\label{subsec:attacks}

Tampering with LLM watermarks can be done by two high-level operations.
\emph{Rephrasing} rewrites the whole text through global LLM paraphrasing or back-translation, whereas \emph{word edits} apply local lexical substitutions, typically via BERT-based attacks \citep{li2020bert,garg2020bae}.
Both are traditionally studied in the no-box setting: prior work scores candidates with proxies of the watermark signal (self-information \citep{cheng2025revealing}, green-list frequency \citep{chen2025mark}, or learned corpora~\citep{jovanovic2024watermark}), because exact token scores were private.
A public detector provides exact feedback that lets an informed attacker alter the watermark score directly.
We quantify the vulnerability this access adds by extending previous attacks to the informed setting, in the white-box and the black-box tiers\footnote{We also considered the tool \href{https://github.com/guillaumemeyer/watermarks-remover}{github.com/guillaumemeyer/watermarks-remover} which removes watermarks by character-level normalization then rephrasing with proprietary LLMs.
The former is unnecessary, while the latter is equivalent to our rephrasing attack run through a closed API.
}.

\paragraph{Removal.}
\begin{itemize}[leftmargin=1.2em, labelsep=0.5em, itemsep=0.1em]
    \item \textbf{Rephrasing}:
    In the no-box setting, an LLM paraphrases the text.
    In the black-box setting, the attacker tries paraphrases at increasing sampling temperatures and keeps the first one the detector no longer flags. The tuned parameter is then the maximum temperature it will try, and the query cost is the number of paraphrases submitted.
    In the white-box setting, the detector allows for regenerating the text with sampling explicitly steered away from the watermark.
    We use two mechanisms: 
    (i)~\emph{Gumbel-max} {\footnotesize (Gumbel)} applies the Gumbel trick with the opposite $r$, thus $x_t=\arg\max_v (1-r_v)^{1/p_v}$, so tokens with low watermark score are favored 
    (ii)~\emph{Additive} {\footnotesize (Add.)} modifies the temperature-scaled logits before multinomial sampling, $\ell_v \gets \ell_v - \lambda \log r_v$ with strength $\lambda>0$, similar to the tunable score-based logit bias of~\citet{jovanovic2024watermark, chen2025mark}.
    In both cases, the sampling temperature is used to control rephrasing quality.
    \item \textbf{Word edits}: In the no-box setting, a ratio of tokens is randomly selected, expanded to a full word (avoiding subword artifacts), and masked.
    A masked language model (MLM) then creates a list of plausible replacement candidates for it, and the most likely replacement is chosen.
    In the black-box setting, replacements are chosen one word at a time and the detector is queried after each edit, and the attack stops at the first edit the detector no longer flags, or terminates after reaching the edit ratio.
    In the white-box setting, at each step the highest-scoring token is identified and replaced by the lowest-scoring of the MLM candidates.
    Edits repeat until the ratio of substitutions is reached, matching the quality degradation of the no-box setting.
    Prompts, hyperparameters and the full procedure (Alg.~\ref{alg:inplace}) are in App.~\ref{app:attacks}.
\end{itemize}

\paragraph{Forgery.}
Existing approaches learn a reverse-engineered watermark from a corpus and forge it onto carrier text \citep{jovanovic2024watermark, chen2025mark}, which is treated in App.~\ref{app:learnability}.
Alternatively, the ``piggyback'' attack maliciously substitutes tokens to change a watermarked text's meaning \citep{pang2024no}, which is identical to the removal workflow.
However, the public detector enables a new type of forgery attack, since its feedback can also be used to raise the score.
We thus adapt the removal attacks using inverted objectives: for rephrasing, we replace $1 - r_v$ by $r_v$ {\footnotesize (Gumbel)} and add $+\lambda \log r_v$ to logits {\footnotesize (Add.)}, while for word edits, we target low-$r$ tokens to pick the highest-scoring replacement.
Black-box forgery via an oracle can be done by iteratively sampling text until a false positive is created~\citep{giboulot2024watermax}.

\section{Split-Key as a Defense Against Informed Attacks}
\label{sec:asymmetric}

\paragraph{Motivation behind the key split.}\label{sec:dichotomy}
A detector that fully exposes the watermark is a liability, as we will show with informed attacks (Sec.~\ref{subsec:vulnerability}). 
This motivates a split construction with two independent keys, one kept private and one made public, each still detecting the watermark on its own.

We score against each key separately: $\mathbf{S}_{\mathrm{pub}}$ and $\mathbf{S}_{\mathrm{priv}}$ are the accumulated per-token scores under $k_{\mathrm{pub}}$ and $k_{\mathrm{priv}}$, each tested against its own null to give two $p$-values $p_{\mathrm{pub}}$ and $p_{\mathrm{priv}}$.
Since routing sends only a fraction $\alpha$ of the tokens to the public key, each channel carries a fraction of the signal.
The public detector returns $p_{\mathrm{pub}}$ to any user, while the platform keeps the detector based on the fused score, which tests $\alpha \mathbf{S}_{\mathrm{pub}} + (1-\alpha)\mathbf{S}_{\mathrm{priv}}$ against the full null for its forensic verdict $p_{\mathrm{fus}}$.

This split benefits the provider on two counts.
First, the private signal is never exposed and is thus much harder to tamper with: an attacker who exploits the public detector steers $p_{\mathrm{pub}}$, the only quantity they observe, which does not determine the platform's verdict on $p_{\mathrm{fus}}$.
Second, the resulting gap between the public and private scores is itself informative, which we exploit in the next paragraph to detect targeted tampering, as shown in Fig.~\ref{fig:threat_model}.
The two keys need not share the same $h$-gram context, and giving them different ones adds a second asymmetry: a longer context is seeded by a wider window, so editing destroys its signal faster, and the gap between the two channels is more sensitive to edits.
\citet{feng2026tracing} proposed such a construction concurrently.
We keep equal contexts throughout the main experiments, and study the unequal variant in App.~\ref{app:asym}.

\paragraph{Tampering test.}\label{subsec:tamper}
We fix $\alpha{=}0.5$ in TextSeal (Sec.~\ref{subsec:related_llmwm}), which makes both channels exchangeable on untampered text (watermarked or not), making per-token differences $d_t = s_{\mathrm{pub},t} - s_{\mathrm{priv},t}$ symmetric around zero. 
An attacker exploiting the public detector disrupts this symmetry, which we measure via $\mathbf{T} = \mathbf{S}_{\mathrm{pub}} - \mathbf{S}_{\mathrm{priv}} = \sum_{t=1}^n d_t$: removal decreases public signal ($\mathbf{T} < 0$), whereas forgery increases it ($\mathbf{T} > 0$).
Both tails test the same null $H_0$, that the keys are exchangeable, 
against $\mathbf{T}<0$ for removal and $\mathbf{T}>0$ for forgery.
Because $\mathbf{T}$'s distribution lacks a closed form under watermarking, we calibrate this null by an exact sign-flip permutation test~\citep{ernst2004permutation,lehmann2022testing}, drawing $R$ random sign vectors $\boldsymbol{\varepsilon}^{(r)} \in \{-1,+1\}^n$ to compute $\mathbf{T}^{(r)} = \sum_{t=1}^n \varepsilon_t^{(r)} d_t$.
The resulting $p$-values (with granularity controlled by $R$, see App.~\ref{app:tampering_details}) are:
\begin{equation}
  p_{\mathrm{removal}} = \frac{1 + \sum_{r=1}^R \mathbbm{1}[\mathbf{T}^{(r)} \le \mathbf{T}]}{1+R}, 
  \quad
  p_{\mathrm{forgery}} = \frac{1 + \sum_{r=1}^R \mathbbm{1}[\mathbf{T}^{(r)} \ge \mathbf{T}]}{1+R}.
  \label{eq:tamper_pvals}
\end{equation}

\begin{figure}[b!]
\centering
\includegraphics[width=\linewidth, clip, trim={0 7.3in 4.4in 0}]{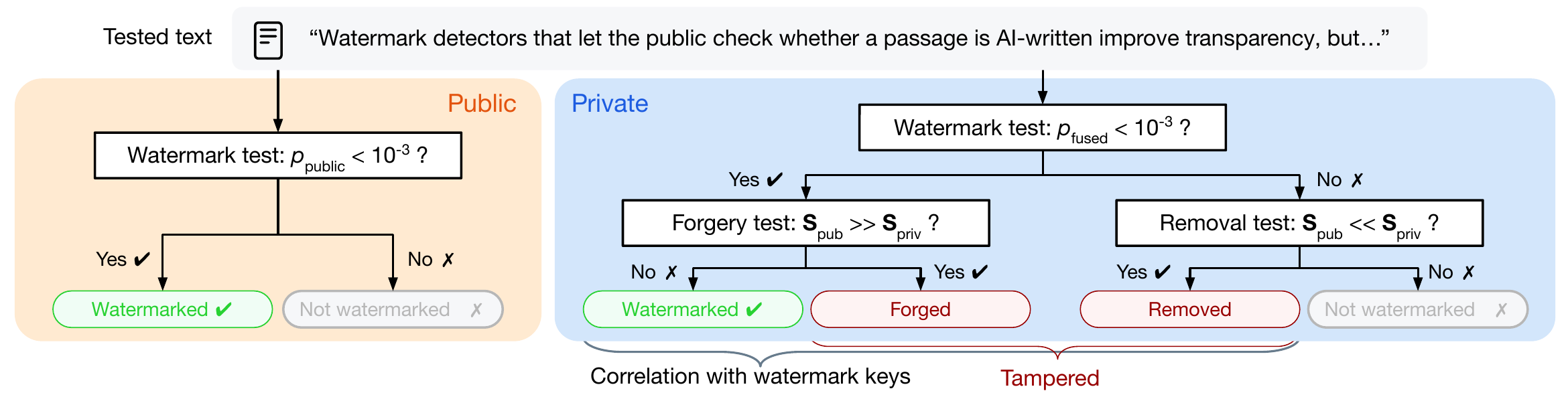}
\vspace{-1.5em}
\caption{%
\textbf{Overview of the two-stage test.}
\emph{(Left)} The public detector only tests based on the public $p$-value $p_{\text{pub}}$. 
\emph{(Right)} The private detector first tests $p_{\text{fus}}$ to get $y_{\text{fus}}$, then re-evaluates it with the tampering test on $\mathbf{T}=\mathbf{S}_{\text{pub}}-\mathbf{S}_{\text{priv}}$. 
If $y_{\text{fus}}=1$ we test for forgery ($s_{\text{pub}} \gg s_{\text{priv}}$), 
if $y_{\text{fus}}=0$ for removal ($s_{\text{pub}} \ll s_{\text{priv}}$). 
Genuine watermarked texts have $\mathbb{E}[\mathbf{T}]=0$, while tampered ones are flagged as forged or removed.
}
\label{fig:attribution_tamper}
\end{figure}

\paragraph{Two-stage mechanism.}\label{subsec:two_stage} 
We construct a two-stage testing mechanism, consisting of the watermark test on the fused score, followed by the tampering test, similar to the image watermarking construction of~\citet{evennou2026forensic}.
The first stage compares $p_{\mathrm{fus}}$ to the threshold set by the target False Positive Rate (FPR), which decides the binary watermark label $y_{\mathrm{fus}}$.
The second stage re-evaluates that watermark verdict with the tampering test.
A text not detected as watermarked is tested for removal, while one detected as watermarked is tested for forgery.
The two stages test different nulls: %
$p_{\mathrm{fus}}$ that the text carries no watermark, and $p_{\mathrm{removal}}$ or $p_{\mathrm{forgery}}$ that its two channels are balanced.
Selecting the tail with $y_{\mathrm{fus}}$ leaves both tails calibrated, since relabeling the channels alters their difference but not their sum which determines $y_{\mathrm{fus}}$ (App.~\ref{app:tampering_details}).
Unwatermarked text can be failed by either test, so its FPR is $2\times10^{-3}$, while a genuine watermarked text faces only the forgery tail, and $\mathrm{ADR}$ and $\mathrm{ASR}_{\mathrm{net}}$ are conditional on the first-stage verdict, each at $10^{-3}$.
An attack informed by the public detector is subject to a tradeoff it cannot observe: tampering with $p_{\mathrm{pub}}$ enough to modify $p_{\mathrm{fus}}$ also enlarges $|\mathbf{T}|$, and is thus caught by the second stage.
The process is depicted in Fig.~\ref{fig:attribution_tamper}. %

\section{Experiments}
\label{sec:experiments}

After setup (Sec.~\ref{subsec:setup}), we measure transparent detector risks (Sec.~\ref{subsec:vulnerability}), evaluate the split-key defense against informed attacks (Sec.~\ref{subsec:tamper_results}), test an attacker adapting to the tampering test (Sec.~\ref{sec:tamper_aware}), and look into other watermark types (Sec.~\ref{sec:kgw_synthid}).

\subsection{Experimental setup}
\label{subsec:setup}

\paragraph{Text generation and watermark parameters.}
We generate texts with Qwen2.5-7B~\citep{qwen2024qwen25} unless stated otherwise (we generalize in App.~\ref{app:othermodels}), watermarked with TextSeal using 2 keys, $\alpha{=}0.5$ and symmetric $h{=}3$-gram contexts in each channel.
The core watermarking mechanism is Gumbel-max, with Maryland and SynthID-Text deferred to Sec.~\ref{sec:kgw_synthid}.
We use 1000 samples from ELI5~\citep{fan2019eli5}, a dataset of open-ended questions paired with human-written answers (we extend to a multilingual domain in App.~\ref{app:benchmarks}).
Each dataset entry gives (a) a prompt for the model, used to obtain watermarked text for removal attacks, and (b) the human answer as carrier text for forgery attacks.
We set the maximum generation length to 400 tokens (effectively achieving an average of 360 tokens), and default $\mathrm{top}_p{=}0.95$, $\mathrm{top}_k{=}50$ and temperature of $1.0$.
For forgery carriers, we select the 1k longest ELI5 responses, which have 220 tokens on average.
We set the detection threshold for both the watermark test and the tampering test at FPR of $10^{-3}$, and report the watermark strength before any attack in App.~\ref{app:prefix_pvalues}.

\paragraph{Attack parameters.}
We report the white-box and black-box settings of Sec.~\ref{subsec:threat}, and defer gray-box to App.~\ref{app:less_transprent}.
The black-box attacker queries the deployed public detector looking to change its decision:
word edit attacks, all using RoBERTa MLM~\citep{liu2019roberta}, sweep up to the same edit fractions as the other tiers, and rephrasing attacks try temperatures in steps of $0.2$ up to a maximum.
The rephrasing proxy is Qwen2.5-3B at strength $\lambda{=}1$ unless stated otherwise (\textit{cross}-tokenizer rephrasing in App.~\ref{app:crosstok}).

\paragraph{Metrics.}
An attack succeeds when both the watermark detection test and the tampering test return the wrong label.
Writing $y$ for the true label, $y_c$ for the label of detector $c \in \{\mathrm{pub}, \mathrm{fus}\}$, and $t=1$ the tampering flag event, we define from the two stages of Sec.~\ref{subsec:two_stage}, the Attack Success Rate (ASR) on the detector $c$, or $\mathrm{net}$ after the joint tampering test, and attack detection rates (ADR) as:
\begin{equation}
\mathrm{ASR}_c = \Pr[y_c \neq y],
\quad
\mathrm{ADR} = \Pr[t = 1 \mid y_{\mathrm{fus}} \neq y],
\quad
\mathrm{ASR}_{\mathrm{net}} = \mathrm{ASR}_{\mathrm{fus}}\,(1 - \mathrm{ADR}).
\label{eq:metrics}
\end{equation}
$\mathrm{ASR}_{\mathrm{pub}}$ is the success the attacker observes, $\mathrm{ASR}_{\mathrm{fus}}$ the one the defender sees before the second stage, and $\mathrm{ASR}_{\mathrm{net}}$ what survives both stages.
Where useful, we also report the watermark signal directly as $-\log_{10} p$, where $p$ is the $p$-value returned by the watermark detection test, so a larger value means stronger evidence of a watermark.
We measure the text \textit{distortion} caused by the attack as $1 - \text{BERT-F1}$~\citep{zhang2020bertscore} against the pre-attack text, which is $0$ for untouched texts and grows with the amount of editing.

\subsection{Vulnerability of Public Detection}
\label{subsec:vulnerability}

\begin{table}[tb]
\centering
\captionsetup{font=small}
\caption{
Removal and forgery ASR at three access levels, on baselines with $K$ keys, where every key is published, for Gumbel-max and Maryland ($\gamma{=}0.25,~\delta{=}2$) watermarks.
Attacks are additive rephrasing and $10\%$ word edits, while distortion is $1-$BERT-F1 ($\downarrow$), pooled over the four watermarks.
A white-box detector is the most exploitable tier, and the only setting that directly enables forgery.
Using a black-box detector as a stopping rule gives removal a few points of ASR at consistently lower distortion.
}
\vspace{-0.8em}
\label{tab:baselines}
\resizebox{\linewidth}{!}{
\setlength{\tabcolsep}{3.0pt}
\begin{tabular}{lccc|ccc|ccc|ccc}
\toprule
 & \multicolumn{6}{c|}{Removal} & \multicolumn{6}{c}{Forgery} \\
\cmidrule(lr){2-7}\cmidrule(l){8-13}
 & \multicolumn{3}{c|}{Rephrase \footnotesize (Add.)} & \multicolumn{3}{c|}{Word edits \footnotesize (10\%)} & \multicolumn{3}{c|}{Rephrase \footnotesize (Add.)} & \multicolumn{3}{c}{Word edits \footnotesize (10\%)} \\
\cmidrule(lr){2-4}\cmidrule(lr){5-7}\cmidrule(lr){8-10}\cmidrule(lr){11-13}
Detector: & no-box & black-box & white-box & no-box & black-box & white-box & no-box & black-box & white-box & no-box & black-box & white-box \\
\midrule
Gumbel-max $K{=}1$ & \hc{70.4} & \hc{81.4} & \hc{99.9} & \hc{1.5} & \hc{1.6} & \hc{72.2} & \hc{0.0} & \hc{0.6} & \hc{20.8} & \hc{0.2} & \hc{0.3} & \hc{56.3} \\
Gumbel-max $K{=}2$ & \hc{82.0} & \hc{92.9} & \hc{99.5} & \hc{6.7} & \hc{8.4} & \hc{97.8} & \hc{0.1} & \hc{0.3} & \hc{6.2} & \hc{0.1} & \hc{0.1} & \hc{48.3} \\
Maryland $K{=}1$ & \hc{90.1} & \hc{96.1} & \hc{100.0} & \hc{8.6} & \hc{10.7} & \hc{15.9} & \hc{0.4} & \hc{1.1} & \hc{78.6} & \hc{0.1} & \hc{0.5} & \hc{1.3} \\
Maryland $K{=}2$ & \hc{95.2} & \hc{98.9} & \hc{100.0} & \hc{34.4} & \hc{40.4} & \hc{98.7} & \hc{0.1} & \hc{0.9} & \hc{38.8} & \hc{0.1} & \hc{0.3} & \hc{1.9} \\
\midrule
Distortion & 0.054 & 0.046 & 0.063 & 0.019 & 0.016 & 0.024 & 0.094 & 0.094 & 0.096 & 0.021 & 0.021 & 0.024 \\
\bottomrule
\end{tabular}

}
\end{table}

We first measure how much a public detector helps an attacker performing the attacks of Sec.~\ref{subsec:attacks}, in the no-box, black-box and white-box settings of Sec.~\ref{subsec:threat}.
Watermarking research assumes that such access trivially enables tampering, which is the reason detectors stay private, yet the size of the advantage has not been systematically quantified.
In Tab.~\ref{tab:baselines} we report the results on the Gumbel-max and Maryland baselines with $K$ keys. 
Gumbel-max with $K=2$ keys is equivalent to TextSeal, while Maryland with $K=2$ keys applies the same routing to two independent green/red lists.

First, the white-box attacks (fully transparent detector) reach far higher success rates than their uninformed counterparts.
This is especially true for forgery, which otherwise requires a corpus (or a valid watermarked sample for the ``piggyback'' case).
Rephrasing reaches about 100\% removal ASR, and forgery reaches up to 79\% ASR on Maryland, which rules out releasing a fully transparent detector.
In the black-box setting (binary result), forgery remains at chance since the detector can only verify if a text has crossed the public detection threshold. 
For removal, the black-box setting improves the ASR-distortion tradeoff.
The attacker tries candidates of increasing strength and keeps the first that flips the verdict, so it changes the text less than the uninformed attack.
This increase costs one query per candidate, which a provider could control with rate limits.

\subsection{Evaluating the Split-Key Defense}
\label{subsec:tamper_results}

\begin{figure}[b!]
\centering
\includegraphics[width=\linewidth]{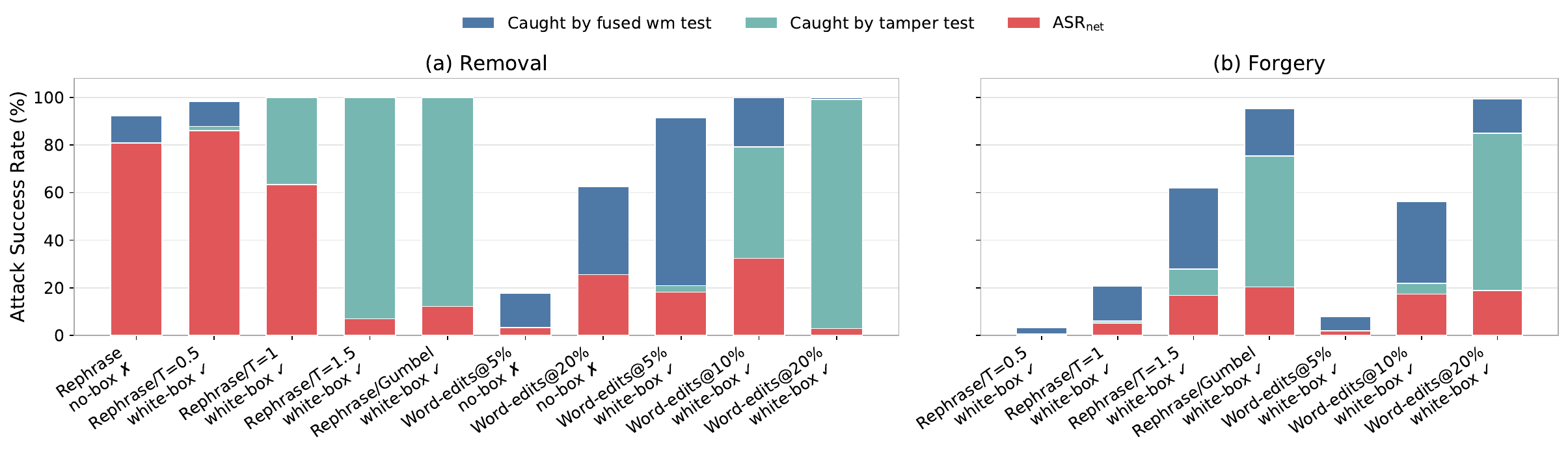}
\vspace{-1em}
\caption{
\textbf{Attack success rate under the two-stage test}, for removal (left) and forgery (right).
Each bar represents the $\mathrm{ASR}_{\mathrm{pub}}$ the attacker sees from the public detector, split into the part the test on the fused watermark score rejects, the part the tampering test then catches, and the $\mathrm{ASR}_{\mathrm{net}}$ that survives both.
Informed attacks fool the public detector, but are overall not more effective than uninformed paraphrasing for removal, and are often flagged for forgery.
}
\label{fig:main_tamper}
\end{figure}

We now run the attacks of Sec.~\ref{subsec:attacks} against the split-key defense of Sec.~\ref{sec:asymmetric}, and report the two-stage effectiveness in Fig.~\ref{fig:main_tamper}.
For removal (left), informed rephrasing and word edits reach near-100\% public ASR, but they widen the public/private gap that the tampering test flags, so fewer attacks are successful after both stages. 
The informed attacker therefore succeeds more often than the no-box one only at the smallest budgets, where it needs fewer edits for the same effect, and less often at the larger ones, where the imbalance it leaves is rejected by the second stage.
Forgery (right) is the opposite case: without access it succeeds only by chance, while public access lets the attacker raise the public scores.
The $\mathrm{ASR}_{\mathrm{net}}$ stays above the near-zero no-box baseline even after the two stages, so public knowledge is useful for forgery.
It is however bounded, since weak attacks do not move the fused score, while strong ones separate the two channels and the tampering test flags most of them.
Full per-attack ASRs and text distortions are in App.~\ref{app:extended_main} (Tab.~\ref{tab:scrub_report} and Tab.~\ref{tab:spoof_report}), while specific text examples before and after the attacks are provided in App.~\ref{app:examples}.

\begin{figure}[b!]
\vspace{-1em}
\centering
\begin{subfigure}[t]{0.53\linewidth}
\centering
\includegraphics[width=\linewidth]{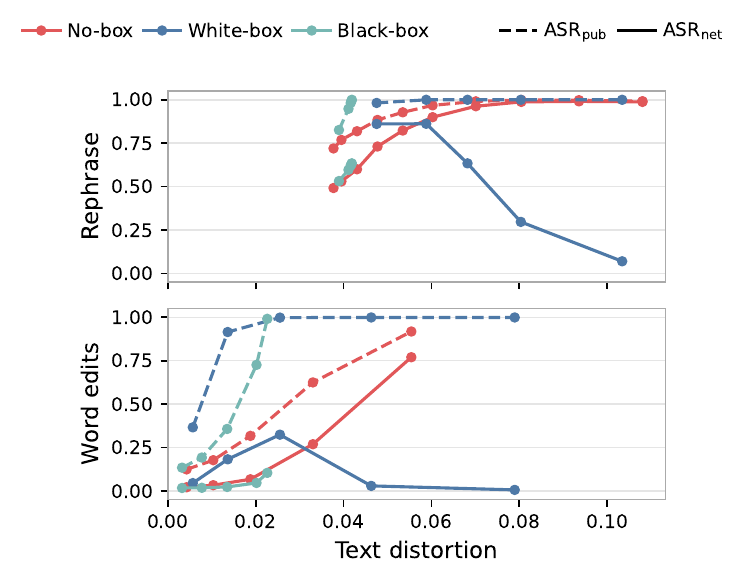}
\vspace{-1.5em}
\caption{Attacker success at each tier.}
\label{fig:asr_distortion_tiers}
\end{subfigure}\hfill
\begin{subfigure}[t]{0.44\linewidth}
\centering
\includegraphics[width=\linewidth]{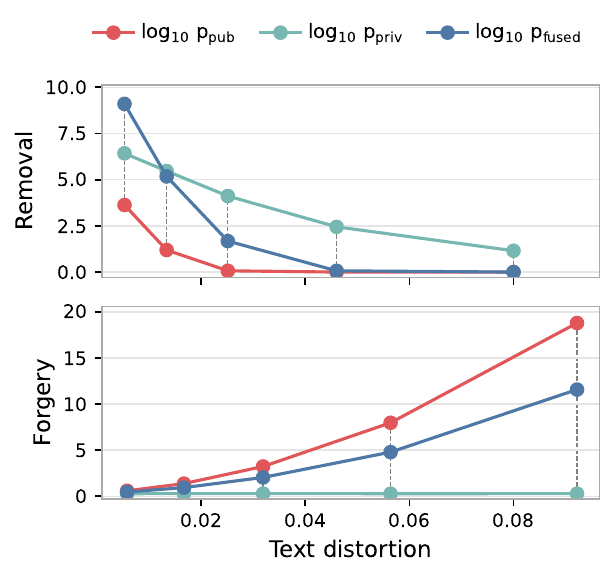}
\vspace{-1.5em}
\caption{%
Signal per channel for white-box attacks.
}
\label{fig:asr_distortion_channels}
\end{subfigure}
\vspace{-0.5em}
\caption{
\emph{(Left)} $\mathrm{ASR}_{\mathrm{pub}}$ (dashed) and $\mathrm{ASR}_{\mathrm{net}}$ (solid) against median distortion, for removal by rephrasing (top) and word edits (bottom), at the \cnobox{no-box}, \cwhitebox{white-box} and \cblackbox{black-box} tiers. The swept parameter is temperature $T \in \{0.5, 0.8, 1.0, 1.2, 1.5\}$ or edit fraction $\{2, 5, 10, 20, 40\}\%$ (for \cblackbox{black-box}, the maximum it will spend). 
The informed attacks with word edits are the most harmful ones for the public detector, but get caught by the two-stage tests (lower $\mathrm{ASR}_{\mathrm{net}}$). 
Overall, informed attacks are stronger only at low distortion, while \cnobox{no-box} rephrasing is more effective at higher distortion.
\emph{(Right)} Median $-\log_{10} p$ of the public, private and fused scores against median distortion, for the white-box word edit sweep, removal (top) and forgery (bottom). 
The private score drops slower than the public one for removal and stays fixed for forgery, so the gap the tampering test examines grows with attack strength.
}
\label{fig:asr_distortion}
\end{figure}

Fig.~\ref{fig:main_tamper} fixes each attack at one operating point, so in Fig.~\ref{fig:asr_distortion_tiers} we sweep removal over rephrasing temperatures and word edit fractions, plotting $\mathrm{ASR}_{\mathrm{pub}}$ and $\mathrm{ASR}_{\mathrm{net}}$ against median distortion.
Informed word edits reach higher public ASR at lower distortion than rephrasing needs for comparable public success, but these stronger attacks are mostly caught by the tampering test: their $\mathrm{ASR}_{\mathrm{net}}$ peaks and then falls below no-box rephrasing.
Fig.~\ref{fig:asr_distortion_channels} shows why, on the informed word edit sweep for removal and forgery: the attacker moves the public score freely (up for forgery, down for removal), the fused score partly follows, and the private score moves slower for removal and stays fixed for forgery.
Therefore, the channel discrepancy the tampering test measures grows with attack strength.
Another view of this tradeoff is to quantify the distortion required to flip the detector (App.~\ref{app:flip_dist}).

The black-box curve in Fig.~\ref{fig:asr_distortion_tiers} behaves differently from both.
With word edits it falls between the other two tiers on the public detector at matched distortion, but comes last on the provider's verdict.
Since it stops the moment the verdict flips, the text ends just past the public boundary, well below the level the fused test needs to reject it.
Rephrasing behaves the same way on the provider's verdict, where it stays at the no-box level, while on the public detector it is the strongest tier at matched distortion.
At this tier no tampering is yet detected (ADR 0\%), because non-targeted edits weaken both channels by the same amount and leave $\mathbf{T}$ unchanged, so $\mathrm{ASR}_{\mathrm{net}}=\mathrm{ASR}_{\mathrm{fus}}$ and the first stage provides the whole defense.
Unequal $h$-gram contexts do not close this gap (App.~\ref{app:asym}).

\begin{table}[t!]
\centering
\captionsetup{font=small}
\caption{
\textbf{Adaptive attacks} against the no-box baseline and their unaware white-box counterparts, at the same operating points (rephrasing $T{=}1$, word edits $20\%$).
Bold marks the highest $\mathrm{ASR}_{\mathrm{net}}$ per family.
Adaptive attacks increase the $\mathrm{ASR}_{\mathrm{net}}$ up to a point, with the private pipeline rejecting a smaller portion of them.
However, they become detectable again at the large overshoot factor of $10^3$.
Despite the limited but real liability of the white-box detector in the adaptive attacks scenario, the $\mathrm{ASR}_{\mathrm{net}}$ with black-box adaptive attacks still matches no-box baselines (App.~\ref{app:blackbox}).
}
\vspace{-0.8em}
\label{tab:adaptive_report}
{\colorlet{heathue}{heatblue!25!white} %
\small
\resizebox{\linewidth}{!}{

\begin{tabular}{l*{14}{C{0.85cm}}}
\toprule
 & \multicolumn{8}{c}{Removal} & \multicolumn{6}{c}{Forgery} \\
\cmidrule(lr){2-9} \cmidrule(lr){10-15}
 & \multicolumn{4}{c}{Rephrase} & \multicolumn{4}{c}{Word edits} & \multicolumn{3}{c}{Rephrase} & \multicolumn{3}{c}{Word edits} \\
\cmidrule(lr){2-5} \cmidrule(lr){6-9} \cmidrule(lr){10-12} \cmidrule(lr){13-15}
Metric & No-box & Adap. ($m{=}10$) & Adap. ($m{=}10^2$) & Adap. ($m{=}10^3$) & No-box & Adap. ($m{=}10$) & Adap. ($m{=}10^2$) & Adap. ($m{=}10^3$) & Adap. ($m{=}10$) & Adap. ($m{=}10^2$) & Adap. ($m{=}10^3$) & Adap. ($m{=}10$) & Adap. ($m{=}10^2$) & Adap. ($m{=}10^3$) \\
\midrule
ASR\textsubscript{pub}$\downarrow$ & \hc{92.3} & \hc{99.5} & \hc{99.9} & \hc{100.0} & \hc{62.5} & \hc{99.9} & \hc{99.8} & \hc{99.8} & \hc{18.1} & \hc{19.6} & \hc{20.8} & \hc{99.8} & \hc{99.9} & \hc{99.9} \\
ASR\textsubscript{fus}$\downarrow$ & \hc{82.0} & \hc{88.5} & \hc{95.1} & \hc{99.4} & \hc{26.9} & \hc{8.8} & \hc{30.8} & \hc{99.3} & \hc{4.8} & \hc{7.5} & \hc{5.5} & \hc{45.8} & \hc{69.8} & \hc{85.1} \\
ADR$\uparrow$ & \cellcolor{heathue!99.6}0.4\% & \cellcolor{heathue!99.8}0.2\% & \cellcolor{heathue!99.8}0.2\% & \cellcolor{heathue!99.3}0.7\% & \cellcolor{heathue!100.0}0.0\% & \cellcolor{heathue!100.0}0.0\% & \cellcolor{heathue!99.7}0.3\% & \cellcolor{heathue!0.6}99.4\% & \cellcolor{heathue!100.0}0.0\% & \cellcolor{heathue!98.7}1.3\% & \cellcolor{heathue!94.5}5.5\% & \cellcolor{heathue!89.1}10.9\% & \cellcolor{heathue!64.0}36.0\% & \cellcolor{heathue!33.8}66.2\% \\
ASR\textsubscript{net}$\downarrow$ & \hc{80.7} & \hc{88.1} & \hc{94.9} & \cellcolor{heathue!98.7}\textbf{98.7\%} & \hc{25.6} & \hc{8.8} & \cellcolor{heathue!30.7}\textbf{30.7\%} & \hc{0.6} & \hc{3.2} & \cellcolor{heathue!5.9}\textbf{5.9\%} & \hc{4.0} & \hc{40.8} & \cellcolor{heathue!44.7}\textbf{44.7\%} & \hc{28.8} \\
\midrule
Distortion$\downarrow$ & 0.052 & 0.052 & 0.053 & 0.057 & 0.032 & 0.012 & 0.016 & 0.060 & 0.096 & 0.096 & 0.096 & 0.040 & 0.046 & 0.050 \\
\bottomrule
\end{tabular}

}
}
\end{table}

\begin{figure}[b!]
\centering
\includegraphics[width=0.83\linewidth]{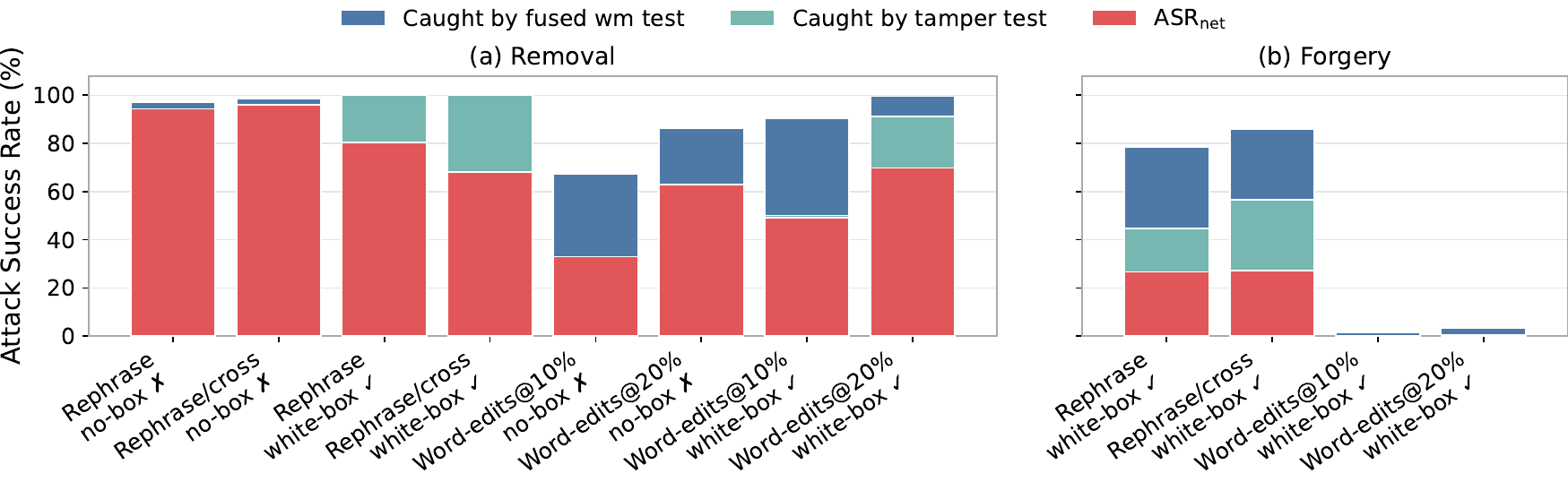}
\includegraphics[width=0.83\linewidth]{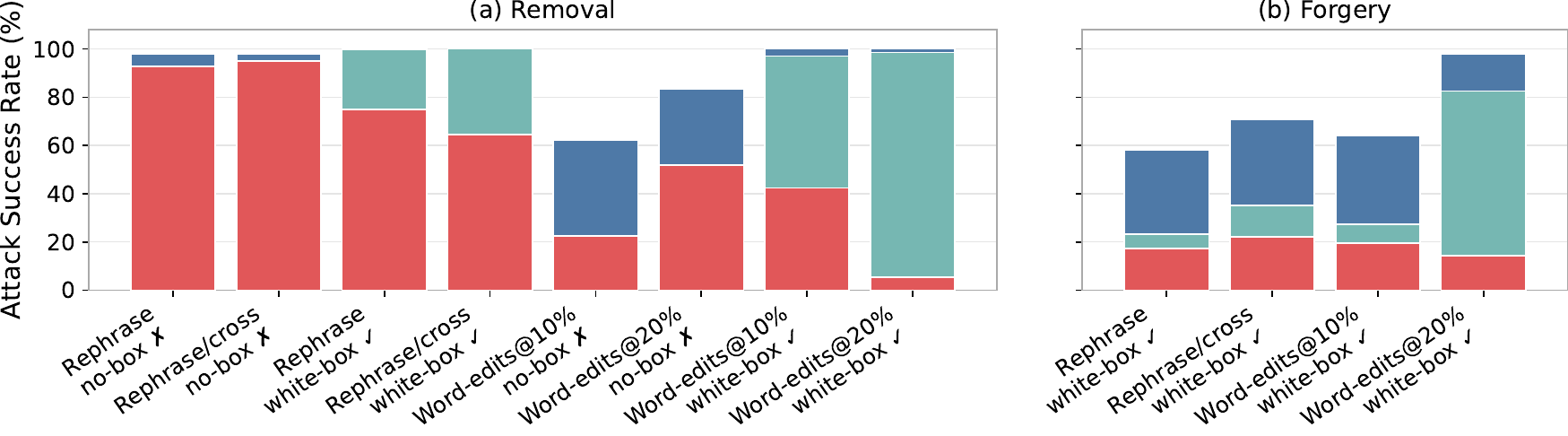}
\vspace{-0.8em}
\caption{%
The split-key construction applies to a Maryland (top) and a SynthID watermark (bottom), where the fused watermark test and the tampering test both reject white-box attacks on the public detector.
Rephrasing is done at temperature of 1 in each case, and we also include cross-tokenizer rephrasing.
White-box rephrasing remains less successful than its no-box counterpart on both schemes, while white-box word edits are more successful at the smaller edit budgets.
}
\label{fig:kgw_synthid}
\end{figure}

\subsection{Adaptive Attacks}
\label{sec:tamper_aware}

So far, we assumed attackers who only target the public detector and ignore the two-stage defense. 
We now consider a more careful, ``adaptive'' attacker.
As noted in Sec.~\ref{subsec:two_stage}, stopping at the detection boundary leaves enough private signal for the fused test, while pushing further widens the gap the tampering test identifies. 
The adaptive attack aims between the two and overshoots the stopping $p$-value by a factor $m$, changing the target to $p_\mathrm{target} {\cdot} m$ for removal and $p_\mathrm{target} / m$ for forgery. 
For word edits, the attack stops once the text reaches the target. 
For additive rephrasing, we scale the bias with the distance to the target and replace the fixed removal bias $\lambda$ by $\lambda \log \left( p_{\text{target}} / p_{\text{current}} \right)$ clipped to $[0, \lambda]$ (inverted ratio for forgery, App.~\ref{app:attacks}), where $p_{\text{current}}$ is $p_\mathrm{pub}$ of the text generated so far. 
For white-box, Tab.~\ref{tab:adaptive_report} compares adaptive to unaware attacks at the same operating points. 
It shows that adaptive attacks evade the tampering test (low ADR), which raises net success against the unaware version. 
For rephrasing, removal slightly exceeds the no-box paraphrase at the same distortion, and it stays low for edits, while forgery stays low for rephrasing and rises for edits but remains partial.

The black-box attacks only observe the yes/no verdict, and stop just at the boundary.
This usually gets flagged by the more powerful fused test.
The adaptive attacker must thus keep editing after crossing the threshold. 
Since this tier returns no score to aim at, the same overshoot factor $m$ is applied in edits instead of in $p$-value: if the verdict flips after $e$ edits, the attacker continues to $m {\cdot} e$ edits, up to the full budget. 
$\mathrm{ASR}_{\mathrm{net}}$ increases with $m$ and remains below the no-box attack. At matched distortion, a partial overshoot reaches a slightly higher $\mathrm{ASR}_{\mathrm{net}}$ than no-box (App.~\ref{app:blackbox}).

Offline attacks that learn the watermark from a corpus are evaluated in App.~\ref{app:learnability}.
They need no detector at all, and forge most carrier texts without moving the two channels apart, but only at $h{=}2$: at the $h{=}3$ used here they reach at most $3\%$ ASR, so this threat is dependent on the context length.

\subsection{Extension to Other Schemes}
\label{sec:kgw_synthid}

So far we have focused on TextSeal's Gumbel-max core, but dual-key routing extends to other watermarks.
We implement Maryland \citep{kirchenbauer2023watermark} and SynthID-Text \citep{dathathri2024scalable} variants under the same construction of routing each position to the public key with probability $\alpha{=}0.5$, and to the private key otherwise.
In the Maryland variant, each key defines its own green list, and generation adds $\delta$ to the green tokens of the routed key.
Similarly, with SynthID-Text each key defines a per-round partition, and the per-token score is the tournament total across rounds.
In both cases, detection is a Binomial tail for both the public and fused detectors, and the tampering test remains the score difference against a channel permutation null.      

The Maryland parameters are $\gamma{=}0.25$ and $\delta{=}2$, and SynthID-Text uses tournament depth $30$ at $\gamma{=}0.5$, while both use the same context of $h{=}3$ tokens per channel.
Both use the same dataset, model, and generation settings as the main TextSeal runs.
The effectiveness of the split construction is presented in Fig.~\ref{fig:kgw_synthid}, following Fig.~\ref{fig:main_tamper}.
ADR is high for white-box attacks on both goals, showing that tampering using the public detector does not help the attacker and raises the chance of detection.

\section{Related Work}
\label{sec:related_short}

\paragraph{LLM watermarking} was introduced by \citet{aaronson2022}, with the non-distortionary Gumbel-max method using pseudorandom next-token sampling, and \citet{kirchenbauer2023watermark}, whose distortionary method boosts a green-list of logits (follow-ups achieved unbiased distributions by scaling this bias~\citep{hu2024unbiased}).
Other non-distortionary methods include inverse sampling \citep{kuditipudirobust}, 
SynthID-Text~\citep{dathathri2024scalable}, 
Permute-and-Flip~\citep{zhao2024permute}, 
TextSeal~\citep{textseal2026}, etc.
More details and methods are given in App.~\ref{app:llm_watermarking}.

\paragraph{Public detectability.}
\citet{christ2024pseudorandom} introduced cryptography-based watermarking, later extended by \citet{fairoze2025publicly} via payload-embedded cryptographic signatures.
Alternatively, public security can rely on independent embedding and detection networks \citep{liu2024unforgeable}, ensuring unforgeability even with publicly shared detectors.
Concurrent work by \citet{feng2026tracing} co-embeds robust and fragile watermarks with independent keys to detect uninformed tampering.%

\paragraph{Removal and forgery.}
Corpus-stealing attacks reverse-engineer the watermark from collected outputs, from Maryland with known PRFs~\citep{jovanovic2024watermark} to any PRF~\citep{chen2025mark} and any logit reweighting~\citep{zhang2026beyond}.
Some works also fine-tune paraphrasers for these tasks: \citet{huang2026rlspoofer} fine-tune a rephrasing model towards forgery, while \citet{diaa2025optimizing} use preference optimization against known watermarking methods and greatly improve no-box removal.
These attacks focus on the no-box scenario, which supports our claim since our baselines use plain paraphrasing or word-editing, which are weaker yet already close to the informed attacks.

\section{Conclusion}

This work examines to what extent the release of a public LLM watermark detector is a liability to the LLM provider.
Given that no-box attacks are possible in any watermark deployment, we investigate whether a public detector can enable stronger watermark tampering.
We build informed attacks that exploit a public detector carrying half of the watermark signal, and manipulate its detection result.
However, the private half of the watermark can reveal the deviation in tampered texts, which we quantify in a statistical test for informed tampering.
The split construction and the tampering test lower the net success rate of informed attacks, although not as effective against the careful attacker who keeps the channel imbalance small.
Broadly, we show that in the case of watermark detection, a split construction can be used as a practical compromise between transparency and security, since it leaves removal close to what the no-detector setting already allows, and keeps the forgery it enables mostly detectable.
Watermark detection could therefore be public: even the strongest attacks that a transparent detector enables leave the provider with a liability that is real but bounded by the key split.
App.~\ref{app:limitations} discusses the scope and limitations of the construction.

\clearpage

\subsection*{Ethics Statement}

This work studies whether LLM watermark detection can be public.
Regulations require AI-generated content to be marked and detectable~\citep{EuropeanAIAct,CaliforniaAITransparencyAct}, yet most providers keep their detectors private because an exposed detector could help attackers.
To make informed decisions on public detection, policymakers and providers need to know how large this risk is, which requires building and evaluating such attacks.
We introduce informed removal and forgery attacks that exploit a public detector, which an actor could use to hide LLM use or to falsely attribute a text to a provider.
These attacks need query access to a detector, so they do not apply to current deployments that keep it private.
We run them only against our own watermark implementations on open-weight models.

We also introduce a way to publish detection results: the provider releases one of two watermark keys through a public detector and keeps the other private.
A two-stage private test on both keys catches most informed forgeries that pass the public detector, including adaptive ones, and informed removal stays close to what is already possible without a detector.

We therefore expect the overall impact of this work to be positive.

\subsection*{Reproducibility Statement}

Experiments build mainly on \url{https://github.com/facebookresearch/textseal}, which has a permissive license (Apache-2.0). 
We will release the experiment code with a permissive license for the camera-ready version.
All models and datasets are publicly available, and no proprietary data is used.
Qwen2.5-7B, SmolLM2, SmolLM3, RoBERTa, MiniLM, GPT-2 and CaLMQA have permissive licenses (Apache-2.0 or MIT).
Qwen2.5-3B and Gemma~3 have custom licenses that allow research use (Qwen Research License, Gemma Terms of Use), and ELI5 consists of public Reddit posts.

\bibliographystyle{ieeenat_fullname}
\bibliography{cite}

\clearpage
\beginappendix
\section{Discussion and Limitations}
\label{app:limitations}

\paragraph{On the channel split.}
A white-box public detector could in principle leak the channel routing: on watermarked text, tokens routed to the private key score low on the public detector, so an attacker could tell the two channels apart.
For removal, using this routing means also editing the low-scoring tokens, but those edits get no detector guidance and amount to uninformed editing, which our no-box baselines already cover.
For forgery there is nothing to leak, since non-watermarked carrier text scores are uniform on each key.
In any case, a realistic deployment would be black-box and return no per-token scores; it can be further hardened with rate limits or a randomized band around the threshold that answers by coin flip, as done against model extraction~\citep{zheng2019bdpl}.
We leave these defenses to future work and evaluate the fully transparent case as a liability upper bound.

\paragraph{On the value of a public detector.}
One objection is that a public detector is pointless, since no-box attacks are already effective.
However, there are multiple reasons in favor of the release of a public detector.
First, verification is no longer proprietary: anyone can examine a text for free without a prior relationship with the provider, which serves the interoperability that Article 50 requires of marking solutions, as far as is technically feasible~\citep{EuropeanAIAct}.
Second, agentic tools are becoming increasingly popular, and produce large portions of LLM-generated text which may be shared without any tampering, so the public detector is still an accessible tool helping against misinformation and fraud. For instance, a user may wish to verify whether a received email has been produced by a personal agent, without relying on stylistic cues.

Regarding exploitation risk, anyone can repeatedly query the detector and edit their text against the returned result, which is the only signal most entities have access to.
However, we have shown that such exploitation is itself detectable: steering with the detector's feedback moves the public channel away from the private one, which is identified in the full detection pipeline.
The latter can still be licensed to trusted partners such as social media platforms, regulatory or investigative bodies, and plagiarism screening tools.
This should deter the informed attack, as it leaves the attacker with rephrasing only, which is an already existing risk, and raises questions on the responsibility of the provider for marking a text that is no longer its original version\footnote{The philosophical paradox of the Ship of Theseus: does a ship fundamentally remain the same object after its original parts are replaced over time?}.

\paragraph{On technical literacy and communication.} 
Non-technical LLM users may misinterpret watermarking and how detection operates.
Even if the provider's public detection tool explicitly states that exploiting it could leave a trace, this will not guarantee rational attack behavior.
However, an intuitive public detection platform, with appropriate information for the user, would still be important towards achieving transparency.

\FloatBarrier

\section{Extended Related Work}
\label{app:related_extended}

\subsection{LLM watermarking}
\label{app:llm_watermarking}

Early work such as \citet{topkara2005natural,topkara2006hiding} performed character-level, syntactic, or synonym substitutions for post-hoc text watermarking.
Such methods have low capacity and robustness, now superseded by generation-time techniques that are compatible with autoregressive LLMs.

A recent research effort has been put into in-generation statistical watermarks.
\citet{aaronson2022} pioneered the non-distortionary Gumbel-max method which samples via a pseudorandom rule instead of multinomial sampling for the next token selection.
The Maryland method \citep{kirchenbauer2023watermark}, although distortionary, is a popular reference method since it boosts a portion of the logits at each generation step.
Follow-up methods such as \citet{hu2024unbiased,wudipmark} scaled the Maryland logit bias with the LLM distribution, achieving a non-distorted next-token distribution.
Other non-distorting methods include inverse sampling \citep{kuditipudirobust}, and SynthID-Text \citep{dathathri2024scalable}, which is based on multilayer tournament sampling.
Permute-and-Flip decoding \citep{zhao2024permute} permutes the candidates with secret randomness and introduces a Gumbel-like watermark test, without changing the output distribution.
OpenAI's textGrain \citep{li2026textgrain} is also non-distortionary: it selects a block of a keyed vocabulary partition through optimal transport with Gumbel costs, under a KL budget that limits how much sampling entropy the watermark removes.
We do not use it here, as it was released a few days before the release of this work.
Finally, TextSeal \citep{textseal2026} extends Gumbel-max method by randomly routing to one of two secret keys, balancing detectability and sampling diversity, while also being compatible with speculative decoding where target model cannot use the same key as the draft.
Using multiple keys has also been explored in the context of watermark ensembling, towards increasing the watermark signal \citep{wu2026ensemble}.

Watermarking relies on imperceptibly altering the LLM distribution, thus it works better in high-entropy generations. \citet{tsur2026heavywater} propose Pareto-optimal methods that are detectable even in low-entropy scenarios.
\citet{giboulot2024watermax} propose WaterMax, a best-of-N scheme that uses a watermarking method's randomness to sample text chunks that are both high-quality and strongly detectable.

\subsection{Public detectability and unforgeable watermarks}

\citet{christ2024pseudorandom} introduced robust cryptography-based watermarking. Subsequently, \citet{fairoze2025publicly} proposed a publicly-detectable scheme for language models by embedding cryptographic signatures in the payload, acknowledging the capacity bottleneck of text.
\citet{pmlr-v258-fairoze25a} formalize the requirements for a robust, unforgeable and publicly-detectable watermark, showing existence in principle but intractability in practice without increase in embedding robustness.
Other lines of work maintain security in the public setting without cryptographic guarantees, using independent embedding and detection constructions.
For instance, \citet{liu2024unforgeable} propose a watermark that is embedded and detected by neural networks, and is thus unforgeable even with a publicly shared detector.
\citet{bashir2026asymmetric} propose a publicly detectable watermark based on the Maryland method that asymmetrically biases public and private green-listed tokens.
\citet{aremu2025mitigating} propose the random switching of watermark keys so that genuinely watermarked texts correlate with only one item from the key set, while forged texts produced from a watermarked corpus would correlate with multiple keys.

The concurrent work of \citet{feng2026tracing} co-embeds a robust and a fragile unbiased signal with independent keys, which can indicate tampering under light edits, including piggyback spoofing.
The split score idea was also shared by \citet{shen2026enhancing,li2026dualguard}, who proposed forgery-proof watermarks with two sources of signal.
However, their constructions are distortionary and not strongly detectable, while their split construction is not oriented towards public detection.

\subsection{Removal and forgery}

A recent line of work aims at learning a text watermark purely through an AI platform's outputs \citep{gu2024learnability}.
\citet{jovanovic2024watermark} proposed a practical watermark stealing method, which reverse-engineers a Maryland scheme with a known pseudorandom function (PRF), further generalized to any PRF in De-Mark \citep{chen2025mark}.
A correlation-based test to classify such forged text is proposed by \citet{pmlr-v267-gloaguen25a}.
Similarly, \citet{zhang2026beyond} use an adaptive method to steal any watermark that performs logit reweighting.

\citet{pang2024no} demonstrate that stealing a non-distortionary schema is possible given access to a watermark detector that provides a detection score per sentence.
In practice, they assume a model API that provides top-$k$ completions to considerably reduce the vocabulary search space.
\citet{huang2026rlspoofer} propose RLSpoofer, a forgery paraphraser for biased watermarks trained via reinforcement learning on unwatermarked and watermarked pairs.
\citet{diaa2025optimizing} optimize removal instead: they curate preference pairs with a surrogate key and tune open-weight paraphrasers with DPO against known watermark algorithms, evading all tested watermarks at negligible quality loss without detector queries, and the tuned paraphrasers transfer to unseen watermarks. This extends their optimization-based adaptive attacks on image watermarks \citep{lukas2024leveraging} to text.

\FloatBarrier

\section{Additional Details}

\subsection{Distortion-free watermarking}
\label{app:distortion}

Following~\citet{dathathri2024scalable}, we state the two levels of non-distortion used in Sec.~\ref{subsec:threat}.
Let $p_t$ be the LLM next-token distribution with $p_t(v)$ the mass on candidate $v$, $x_t$ the sampled token, and $r_t(v)$ the seed value for $v$ at position $t$.
Let $\mathcal{V}$ be the vocabulary, $\mathcal{R}$ the seed space, and $\mathcal{K}$ the key space.

\paragraph{Single-token non-distortion} ~\citep[Def.~16]{dathathri2024scalable}.
A sampling rule $A$ mapping $(p_t, r)$ to a token in $\mathcal{V}$ is \emph{single-token non-distortionary} if for any $p_t$ and any $v \in \mathcal{V}$:
$$\mathbb{E}_{r} \left[ \Pr(A(p_t, r) = v) \right] = p_t(v),$$
where the expectation is over a uniform seed $r \in \mathcal{R}$.
Marginalizing over the seed, each token is sampled with exactly its original LLM probability.
This is a property of the sampling rule alone (\eg Gumbel-max or SynthID-Text tournament sampling), independent of how the seed is produced at each step.
In a deployed scheme the seed at $t$ is the PRF output keyed by $k$, so averaging over $k$ matches averaging over $r$ when each context is fresh.

\paragraph{Single-sequence non-distortion} \citep[Def.~20, single-response case]{dathathri2024scalable}.
A watermarking scheme $p_{\mathrm{wm}}$ is \emph{single-sequence non-distortionary} if, for any prompt $c$ and response $x_{1:n} \in \mathcal{V}^n$:
$$\mathbb{E}_{k} \left[ p_{\mathrm{wm}}(x_{1:n} \mid c, k) \right] = p_{\mathrm{LM}}(x_{1:n} \mid c),$$
where $p_{\mathrm{wm}}(x_{1:n} \mid c, k) = \prod_{t=1}^n q_k(x_t \mid c, x_{1:t-1})$, $q_k(\cdot \mid c, x_{1:t-1})$ is the law of $A(p_t, r_t(k))$ with $p_t = p_{\mathrm{LM}}(\cdot \mid c, x_{1:t-1})$, and the expectation is over uniform $k \in \mathcal{K}$.
This is strictly stronger: the \emph{joint} distribution over a full response must match the original LLM, not just the per-token marginals.
A single-token non-distortionary sampler with a sliding-window seed violates it whenever the same context repeats within a generation, since the PRF then returns identical values and re-selects the same token.
\emph{Repeated context masking}~\citep{dathathri2024scalable} restores it by falling back to unwatermarked sampling on repeated contexts.
In practice, TextSeal and SynthID-Text reach answer-level non-distortion through this generation-time dedup: repeated $h$-grams are rare with $h \geq 3$, so masking them changes little while removing the deterministic repetitions.
This scheme has no effect on model output quality and is used in production~\citep{dathathri2024scalable,textseal2026,anthropic2026marking}.

\subsection{Attacks}\label{app:attacks}

We provide additional details for the attack families described in Sec.~\ref{subsec:attacks}.
For the rephrasing attack, in all settings, we use the prompt:

\texttt{f"Rephrase the following text while preserving its meaning. Output only the rephrased version, nothing else.\\
Text: \{text\}\\
Rephrased:"}

Since the word edits attack relies on an MLM for word replacement, which cannot capture math, we explicitly filter out the math-related samples from ELI5 for a fair comparison before isolating the first 1k samples. We detail the word edits attack in Algorithm~\ref{alg:inplace}. The method is similar to the self-information attack of \citet{cheng2025revealing} and the adversarial BERT attacks of \citet{li2020bert} and \citet{garg2020bae}.

\begin{algorithm}
\caption{Word edits for Watermark Removal (Forgery in comments).}
\label{alg:inplace}
\begin{algorithmic}[1]
\Require Text $x$, score threshold $\zeta$, $p$-value goal $\theta$, max edits $B$, max passes
$I$, words per pass $w$
\State $n \gets 0$, ~ $i \gets 0$ \Comment{Edit count, pass count}
\Repeat
\State $p, \{r_v\} \gets \Call{Detector}{x}$
  \Comment{$p$-value and per-token scores (white-box detector)}
\State $i \gets i + 1$
\State $W \gets$ the $w$ words with highest-$r$ tokens
  \Comment{Expanded subword tokens (Forgery: lowest-$r$)}
\For{each word $w \in W$} \Comment{Right to left, keeping earlier scores valid}
    \State $C \gets \Call{MLM}{x, [w]}$ \Comment{Mask $w$ and get a replacement candidate set}
    \For{each candidate $c \in C$}
        \State $x_c \gets x$ with $c$ in place of $w$
        \State $\_, \{r_t(c)\} \gets \Call{Detector}{x_c}$
          \Comment{Score by querying detector}
        \State $\mathbf{S}(c) \gets \sum_{t \in \mathrm{tok}(c)} -\log(1 - r_t(c))$
          \Comment{$\mathrm{tok}(c)$: tokens of $x_c$ containing a part of $c$}
    \EndFor
    \State $C_\zeta \gets \{c \in C : \max_{t \in \mathrm{tok}(c)} r_t(c) < \zeta\}$
      \Comment{Select weak candidates (Forgery: $> \zeta$)}
    \If{$C_\zeta \neq \emptyset$}
        \State Replace $w$ with $\arg\min_{c \in C_\zeta} \mathbf{S}(c)$ in $x$ \Comment{Best candidate
(Forgery: $\arg\max$)}
        \State $n \gets n + 1$
    \EndIf
\EndFor
\Until{$p \geq \theta$ \textbf{or} $n \geq B$ \textbf{or} $i \geq I$}
\Comment{(Forgery: $p < \theta$)}
\State \Return $x$
\end{algorithmic}
\end{algorithm}

\paragraph{Adaptive white-box attacks.}
These are the attacks of Sec.~\ref{sec:tamper_aware}, run at $p_{\text{target}}{=}10^{-3}$ with overshoot factor $m$.
For word edits, Alg.~\ref{alg:inplace} is unchanged except for the stopping test, which becomes $p_\mathrm{pub} \geq m\,p_{\text{target}}$ for removal and $p_\mathrm{pub} < p_{\text{target}}/m$ for forgery. %
For rephrasing, the fixed additive bias $\lambda$ is replaced by $\lambda \min(1, \max(0, \log(p_{\text{target}} / p_{\text{current}})))$ (forgery: $\log(p_{\text{current}} / p_{\text{target}})$), recomputed at every decoding step from $p_{\text{current}}$, the running public $p$-value of the text generated so far.
Both read $p_\mathrm{pub}$, so both need at least the gray-box tier.

\paragraph{Adaptive black-box attacks.}
In this setting, the public detector answers only $\mathbbm{1}[p_\mathrm{pub} < 10^{-3}]$.

For the black-box word edits attack, we first split the text in whitespace-delimited words, and draw uniformly words one by one without replacement.
Replacements come from the same \texttt{roberta-large} MLM as in the white-box attack ($k{=}50$, minimum probability $10^{-4}$). 
One word is replaced at a time and the detector is queried once per edit, so the number of queries equals the edit count. 
The attack stops at the first edit whose verdict flips, or at the edit budget. 
In the adaptive case, the overshoot factor $m$ of Sec.~\ref{sec:tamper_aware} is applied in edits: we keep editing up to $m$ times the number needed for the flip, while staying under the edit budget. 
For instance, if $m=1.5$ and the $10$th edit flipped the decision, we would edit $15$ words if the budget is bigger than $15$ and exactly the budget if the budget is lower than $15$.
$m{=}1$ is the stopping rule from before, and we compare with the no-box setting which spends the whole budget and never uses the verdict (equivalent to a large $m$).

For the black-box rephrasing attack, the attacker increases the temperature $T \in \{0.2, 0.4, \dots, T_{\max}\}$ and submits each paraphrase, keeping the first whose verdict flips.

\subsection{Tampering Test}
\label{app:tampering_details}

\begin{figure}[b!]
\centering
\includegraphics[width=0.9\linewidth]{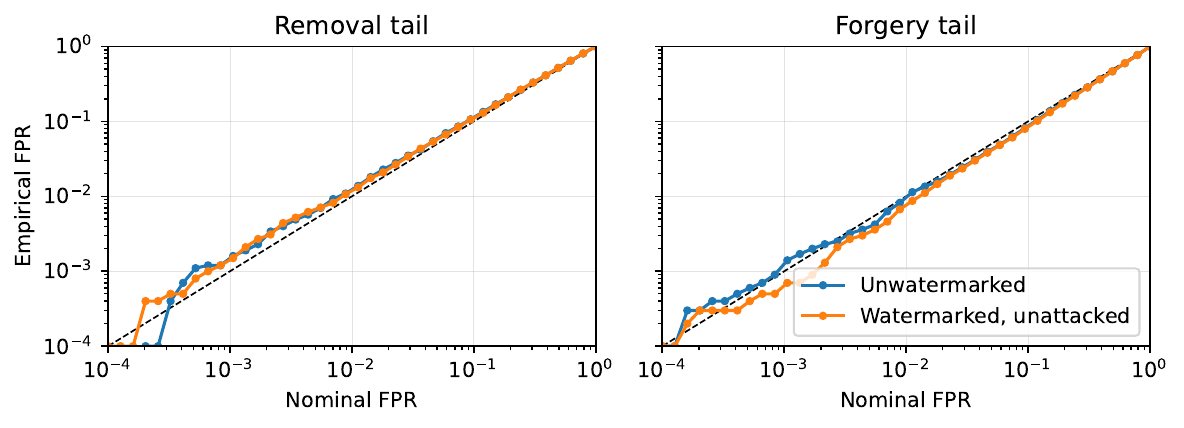}
\includegraphics[width=0.9\linewidth]{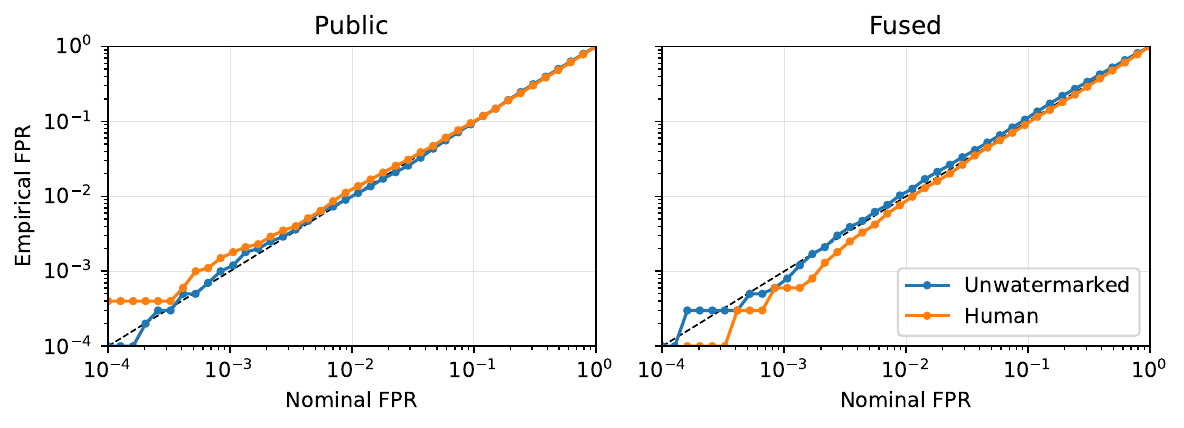}
\caption{\textit{(Top)} Null calibration of the tampering tests on two sets of 10k untampered texts, both generated by Qwen2.5-7B. One is unwatermarked and the other is watermarked but untampered, covering both tampering nulls. \textit{(Bottom)} Null calibration of the watermark tests for both detectors. We use two sets of 10k unwatermarked texts, one generated from Qwen2.5-7B, and the other being the human-written answers in ELI5.}
\label{fig:tamp_null}
\end{figure}

We use $R=5\times10^4$ permutations to compute the $p$-values, which allows for a minimum $p$-value of $2 \times 10^{-5}$, covering the $10^{-3}$ detection threshold.
We also show that the tampering test is calibrated on its nominal FPR.
Fig.~\ref{fig:tamp_null} shows that the empirical FPR closely follows its nominal value, indicating that the tampering test is calibrated on its null assumption of channel symmetry. Additionally, we verify each watermark test's calibration, since we use the fused test in the full detection pipeline, and use only one channel for the public detector.

\paragraph{Rate of the two stages together.}
The plots in Figs.~\ref{fig:tamp_null} are per-stage marginals: each verifies one test against its own null in isolation.
Both corpora in Fig.~\ref{fig:tamp_null} probe the single null $H_0$ of Sec.~\ref{subsec:tamper}, which is why one unwatermarked and one watermarked set both belong there.
Write $\eta$ for the per-stage FPR level, distinct from the routing fraction $\alpha$ of Sec.~\ref{subsec:tamper}. Running the two stages in sequence at a common $\eta$ gives unwatermarked text two chances to be flagged: the first stage may call it watermarked, and failing that the second may call it scrubbed. A genuine watermarked text, whose first-stage verdict is correct, faces only the forgery tail and so only one.
The permutation relabels the channels: swapping the two labels at a position leaves $s_{\mathrm{pub},t}+s_{\mathrm{priv},t}$ unchanged, so the fused score, and with it $p_{\mathrm{fus}}$ and $y_{\mathrm{fus}}$, is constant across every draw of the null.
The tampering $p$-value is therefore exactly uniform conditionally on the first-stage verdict, which is what keeps each stage at its nominal level.
This relies on the first stage reading the sum of the channels, not on it executing first: had it thresholded $p_{\mathrm{pub}}$, selection would favor large $\mathbf{T}$, since $\mathrm{Corr}(\mathbf{S}_{\mathrm{pub}}, \mathbf{S}_{\mathrm{pub}}-\mathbf{S}_{\mathrm{priv}})=1/\sqrt{2}$ at equal variances, and the forgery tail would have increased FPR on untampered text.
The rate at which unwatermarked text draws \emph{any} flag is then $\eta + (1-\eta)\eta \approx 2\eta$ instead of $\eta$.
Tab.~\ref{tab:joint_fpr} measures this on $799$ unwatermarked generations.
The measured union is below $2\eta$ only because the fused watermark test is itself slightly conservative on this corpus.
At $10^{-3}$ the corpus is too small to resolve the rate.

On watermarked untampered text the first stage is correct, so there is only one way to err: a false forgery verdict against a genuine user.
That one stays on nominal, firing on $5.31\%$, $1.05\%$ and $0.17\%$ of the $5992$ untampered generations at $\eta$ of $0.05$, $0.01$ and $10^{-3}$.
The attack rates we report are likewise conditional on a stage-one verdict and so are governed by the single-stage level.
The factor of two applies to the unconditional statement, \ie the rate at which the pipeline flags an arbitrary clean text at all, and splitting the budget as $\eta/2$ per stage restores a joint $\eta$ at a small cost in power.

\begin{table}[t!]
\centering
\caption{Flag rates of the two-stage pipeline on $799$ unwatermarked generations, in percent. The second stage is the removal tail whenever the first stage returns \emph{not watermarked}, and the forgery tail otherwise. Each stage holds its nominal level $\eta$, but unwatermarked text can fail either of them and so is flagged at roughly $2\eta$.}
\label{tab:joint_fpr}
\begin{tabular}{crrrr}
\toprule
$\eta$ & Watermark test & Tampering test & Either stage & $\eta+(1-\eta)\eta$ \\
\midrule
$0.05$ & $3.88$ & $5.01$ & $8.51$ & $9.75$ \\
$0.02$ & $1.13$ & $2.38$ & $3.50$ & $3.96$ \\
$0.01$ & $0.63$ & $0.88$ & $1.50$ & $1.99$ \\
\bottomrule
\end{tabular}

\end{table}

Furthermore, we examine the computational cost of the tampering test compared to the watermark test.
We present the results in Tab.~\ref{tab:timing}.

\begin{table}[t]
\centering
\caption{
Average CPU cost of watermark detection and the tampering test, on the same 1k texts.
Both tests share one fused $K$-key scoring pass (the $43.9$~ms detection cost). The tampering test adds a permutation over $R$ replicates of the already-computed score matrix, so the tampering columns report the permutation alone.
Larger $R$ lowers the smallest resolvable $p$-value but scales the cost linearly. 
We use $R=5\times10^{4}$ by default.}
\label{tab:timing}
\footnotesize
\begin{tabular}{lccccc}
\toprule
 & Detection & \multicolumn{4}{c}{Tampering} \\
\cmidrule(lr){2-2}\cmidrule(lr){3-6}
Permutations $R$ & -- & $5\times10^{3}$ & $5\times10^{4}$ & $2\times10^{5}$ & $1\times10^{6}$ \\
\midrule
Resolvable $\min p$ & analytic & $2\times10^{-4}$ & $2\times10^{-5}$ & $5\times10^{-6}$ & $1\times10^{-6}$ \\
Test duration (ms) & 43.9 & 23.3 & 276.9 & 1105.1 & 5518.5 \\
\bottomrule
\end{tabular}
\end{table}

\paragraph{A permutation instead of a binomial test.}
Exchangeability also admits a binomial test: count the positions with $d_t>0$ against $\mathrm{Binomial}(n,\tfrac12)$, which needs no sampling but discards $|d_t|$.
Neither dominates (Tab.~\ref{tab:signtest}). 
Rephrasing moves nearly every position, so the signs alone carry the evidence and the binomial test is more powerful.
Forgery and word edits concentrate their damage in a few positions, and the permutation doubles the detection rate.
We keep the permutation because its worst case is the better of the two ($1.2\%$ against $0.6\%$) and forgery is the threat the test exists to catch.

\begin{table}[t]
\centering
\caption{Attack detection rate (\%) at $p<10^{-3}$ for the two exact tests, symmetric $(2,2)$, white-box attacks, $1000$ documents per row.
Bold marks the better test per attack.
The binomial is more powerful on rephrasing, the permutation on the concentrated attacks.}
\label{tab:signtest}
\small
\begin{tabular}{lcc}
\toprule
Attack & Binomial & Permutation \\
\midrule
Untampered (FPR) & 0.3 & 0.1 \\
\midrule
Rephrase removal & \textbf{86.9} & 82.4 \\
Rephrase removal (additive) & \textbf{40.1} & 25.1 \\
Word-edit removal & 0.6 & \textbf{1.2} \\
Rephrase forgery & 12.1 & \textbf{25.2} \\
Word-edit forgery & 3.2 & \textbf{9.5} \\
\bottomrule
\end{tabular}

\end{table}

\paragraph{Routing must stay uniform.}
Exchangeability holds because each token goes to either key with probability $\tfrac12$, so relabelling the two is legitimate. Routing them unequally breaks it: a token is then more likely to have come from one key, the sign of $d_t$ is a biased coin, and the sign-flip null no longer applies.

\subsection{Attack Examples}
\label{app:examples}

\begin{figure}[b!]
\centering
\includegraphics[width=\linewidth]{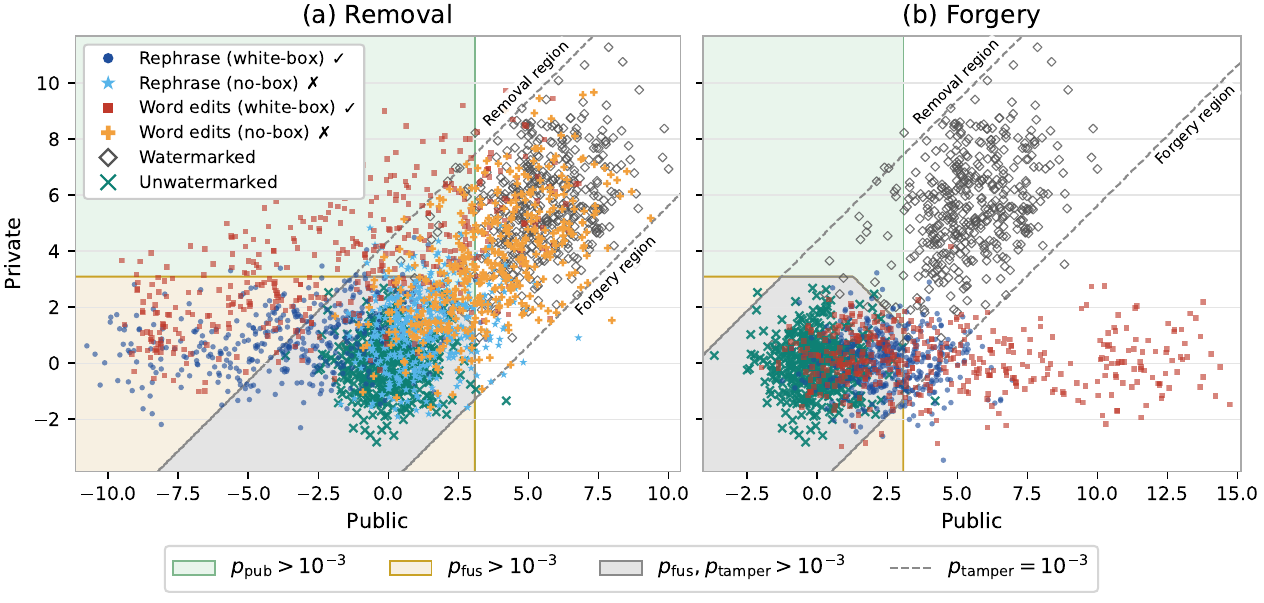}
\caption{
The scores (standardized using their mean and variance) of each channel and the areas where each test fails to reject its null hypothesis (\eg $p_{\mathrm{pub}} > 10^{-3}$ means that the public $p$-value is high enough for the text to be labeled as unwatermarked).
The attacked data points are stratified samples at different strength levels from our main experiments on Qwen2.5-7B (additive rephrasing with a temperature sweep, and word edits with a replacement rate sweep).
The public detector can capture genuinely watermarked text, but is far less sensitive to tampered text than the fused detector and the tampering test (diagonal band, signifying large score deviations).
White-box attacks tend to move the scores outside of the tampering test band, into either the removal or the forgery region, whereas the no-box removal attacks move the scores only within the balanced score band.
The white-box removal attacks also have similar private scores to their no-box counterparts, but are denser towards lower public score areas.
}
\label{fig:pubpriv}
\end{figure}

\begin{figure}[b!]
\centering
{\scriptsize Removal caught by fused:}
\includegraphics[width=\linewidth]{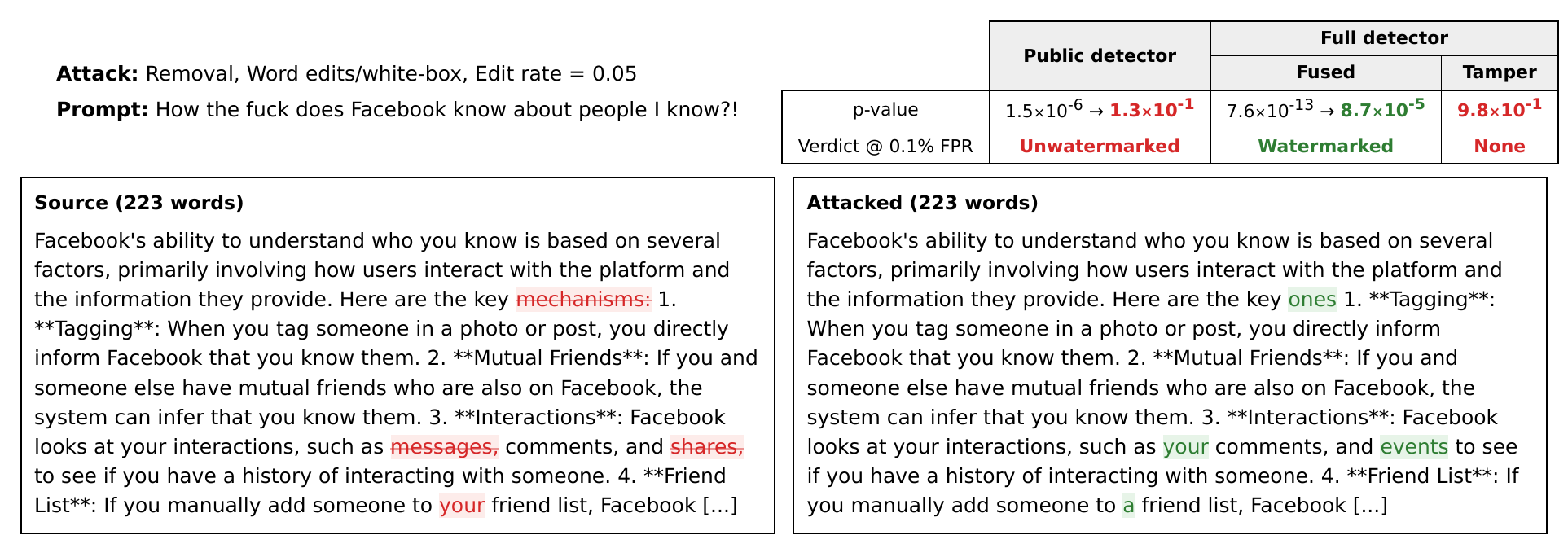}
{\scriptsize Removal caught by tampering:}
\includegraphics[width=\linewidth]{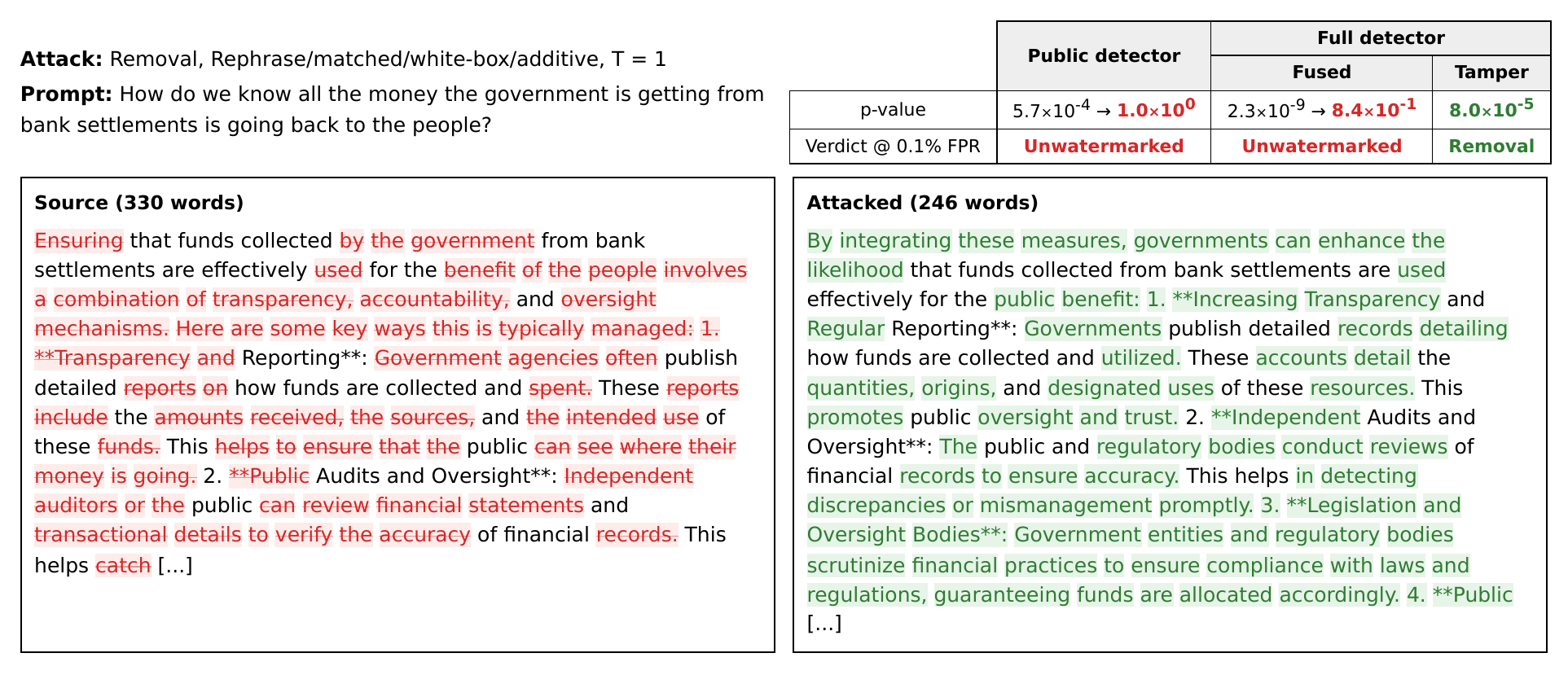}
{\scriptsize Successful removal:}
\includegraphics[width=\linewidth]{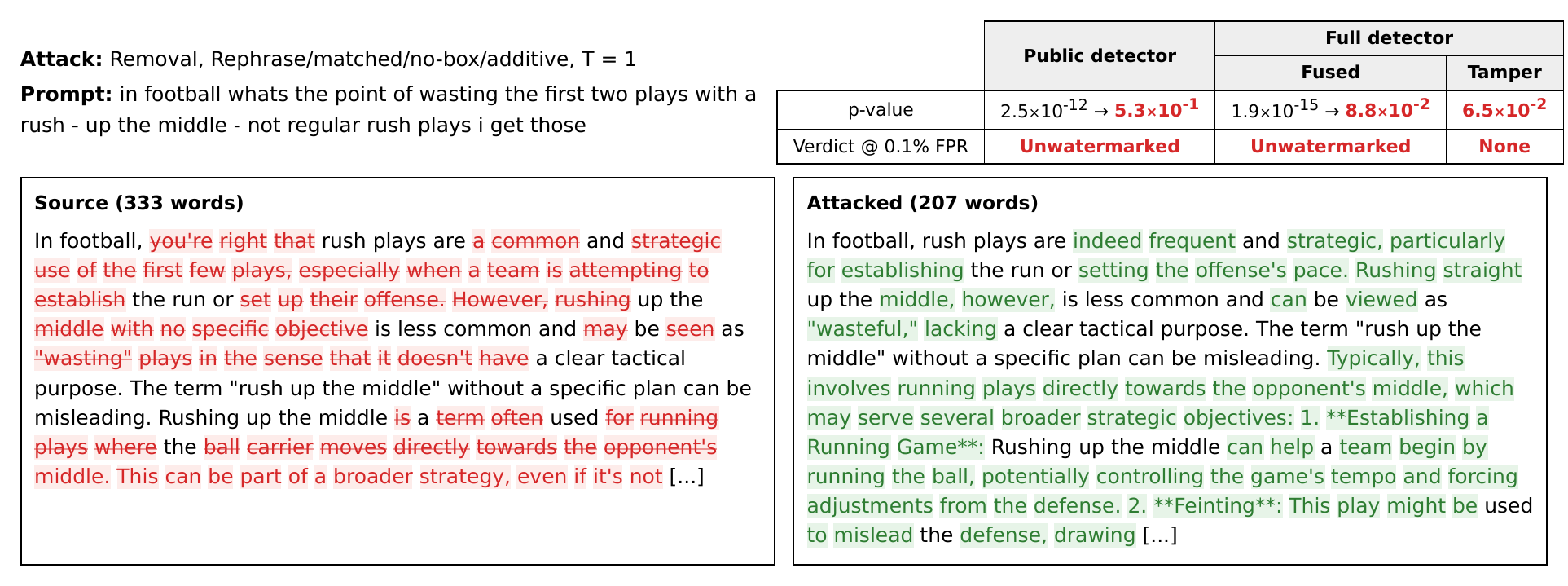}
\caption{
Watermark removal examples with white-box attacks, either rephrasing or word edits.
While the public score may be lowered enough to indicate unwatermarked text, the fused watermark test along with the tampering test can discover that the text is actually either watermarked or tampered, indicating key correlation with the provider's LLM.
}
\label{fig:removal_examples}
\end{figure}

\begin{figure}[b!]
\centering
{\scriptsize Forgery caught by fused:}
\includegraphics[width=\linewidth]{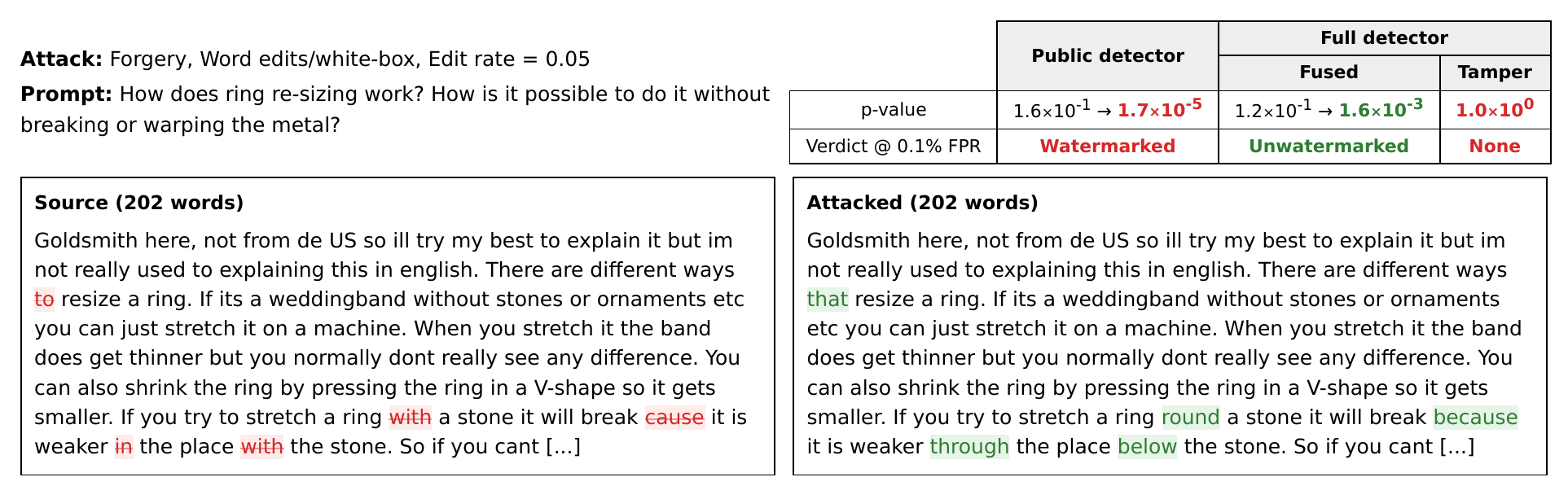}
{\scriptsize Forgery caught by tampering:}
\includegraphics[width=\linewidth]{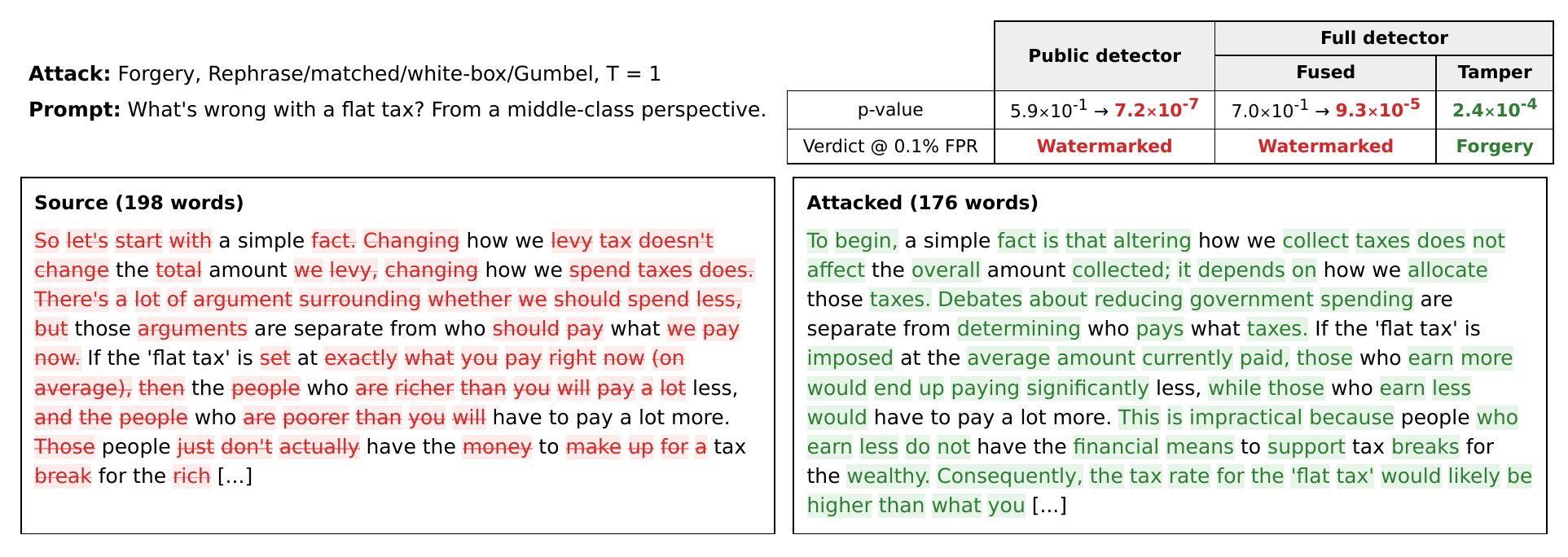}
{\scriptsize Successful forgery:}
\includegraphics[width=\linewidth]{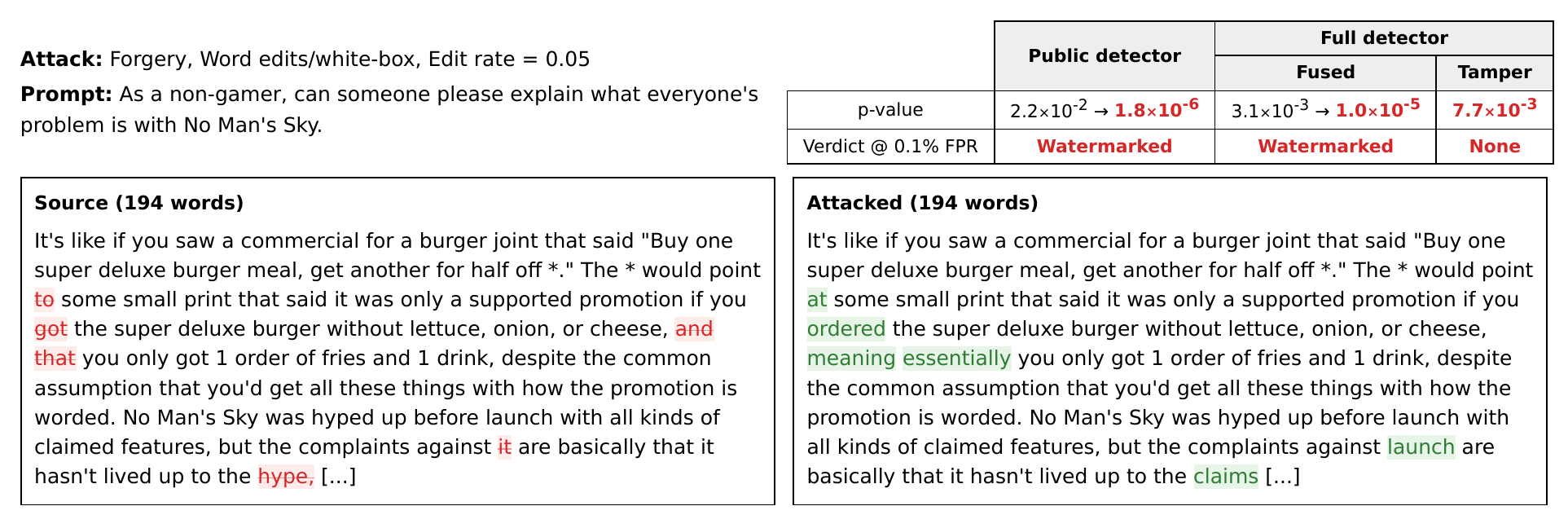}
\caption{
Watermark forgery examples with white-box attacks, either rephrasing or word edits.
While the public score may be increased enough to forge a watermark signal, the full detector is harder to deceive, while the tampering test may explicitly indicate forgery.
}
\label{fig:forgery_examples}
\end{figure}

We visualize how the different attack goals manipulate the watermark score distributions in the score space, and how the tampering test covers different areas than the fused watermark test in Fig.~\ref{fig:pubpriv}. 
We qualitatively show how our white-box attacks manipulate actual texts to remove the watermark in Fig.~\ref{fig:removal_examples}, and to forge the watermark in Fig.~\ref{fig:forgery_examples}.

\section{Additional Results}

\subsection{Extended Numbers of the Main Results}
\label{app:extended_main}

\begin{figure}[b!]
\centering
\includegraphics[width=0.75\linewidth]{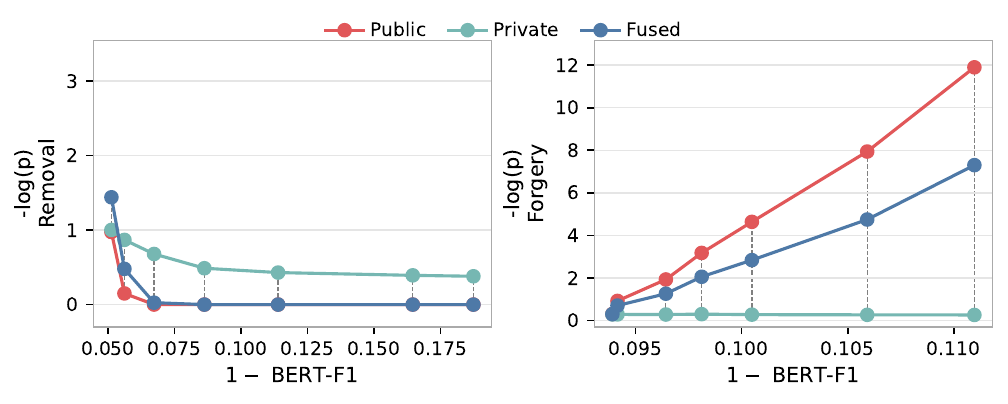}
\caption{
Additive rephrasing with the same model and dataset at $T{=}1$, but with a strength sweep $\lambda \in \{0, 0.5, 1, 1.5, 2, 3, 4\}$. The strength is similarly effective to temperature at manipulating the public score, and the fused score alongside it. However, extreme strengths also cause large channel deviations for forgery, which are detected by the tampering test.
}
\label{fig:pareto_strength}
\end{figure}

Tab.~\ref{tab:scrub_report} and Tab.~\ref{tab:spoof_report} give the per-attack numbers behind Fig.~\ref{fig:main_tamper}, for removal and forgery respectively.
For each attack we report the median $-\log p$ of the public, private, and fused scores, the attacker-visible public success rate $\mathrm{ASR}_{\mathrm{pub}}$, the success rate after the fused-score test $\mathrm{ASR}_{\mathrm{fus}}$, the attack detection rate of the tampering test $\mathrm{ADR}$, the effective threat $\mathrm{ASR}_{\mathrm{net}}$ that survives both stages, and the text distortion $1-$BERT-F1.
We note that the tables report the rates unconditionally.
However, the bar heights in Fig.~\ref{fig:main_tamper} (with a full bar equal to $\mathrm{ASR}_{\mathrm{pub}}$), are conditioned on the public success, since only the texts that fool the public detector are examined.
The two views are almost equivalent and only differ on rare occasions where texts do not evade the public detector but evade the fused one.
From the attacker's perspective, we follow the per-channel score presentation of Fig.~\ref{fig:asr_distortion_channels} and examine the strength $\lambda$ as an additional attack parameter, showing its effect in Fig.~\ref{fig:pareto_strength}.

\begin{figure}[b!]
\centering
\includegraphics[width=\linewidth]{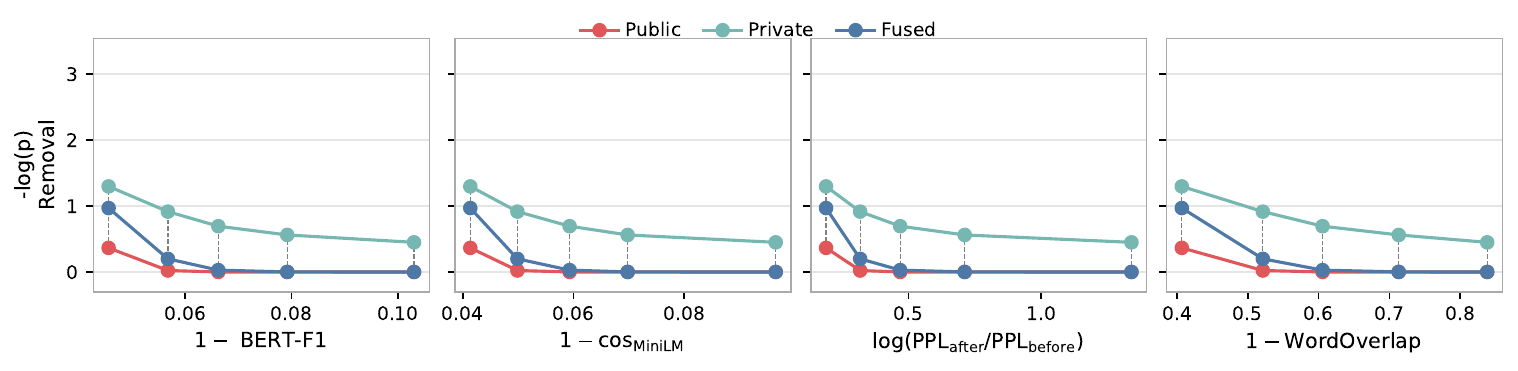}
\includegraphics[width=\linewidth]{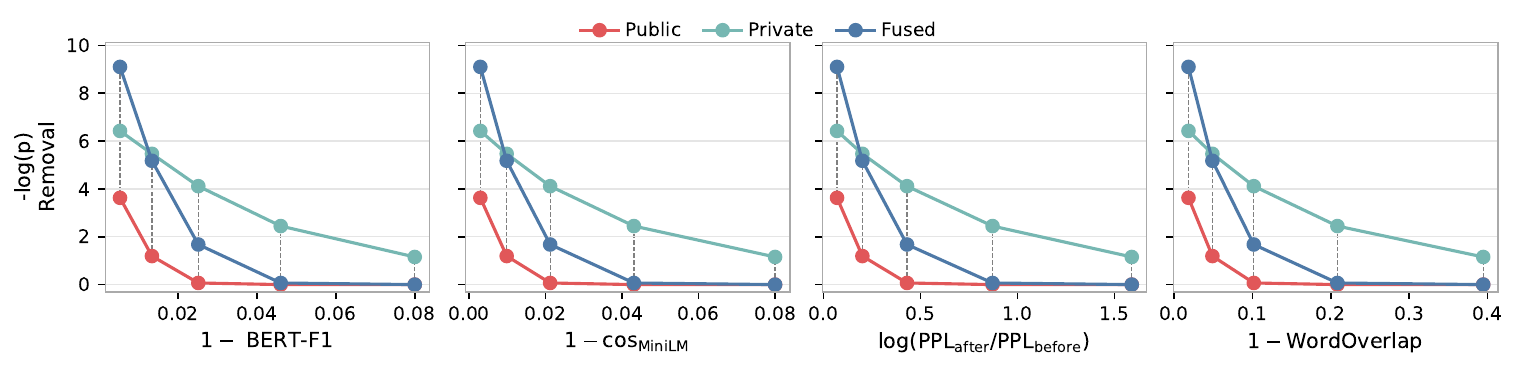}
\caption{
Removal watermark score per channel over quality degradation with the rephrasing attack (top) and word edits (bottom). The settings are the same as Fig.~\ref{fig:asr_distortion}, namely temperature $T \in \{0.5, 0.8, 1.0, 1.2, 1.5\}$ and edit fractions $\{2, 5, 10, 20, 40\}\%$.
The vertical axes are the same across all panels, but the horizontal axes show the different distortion measurements.
Rephrasing causes stronger distortion on pure word overlap compared to word edits (which cause their nominal edit rate), but weaker distortion on fluency as quantified with the perplexity ratio, and similar distortions to the LM metrics (BERT and MiniLM) that capture both.
}
\label{fig:pareto_more_quality_removal}
\end{figure}

\begin{figure}[b!]
\centering
\includegraphics[width=\linewidth]{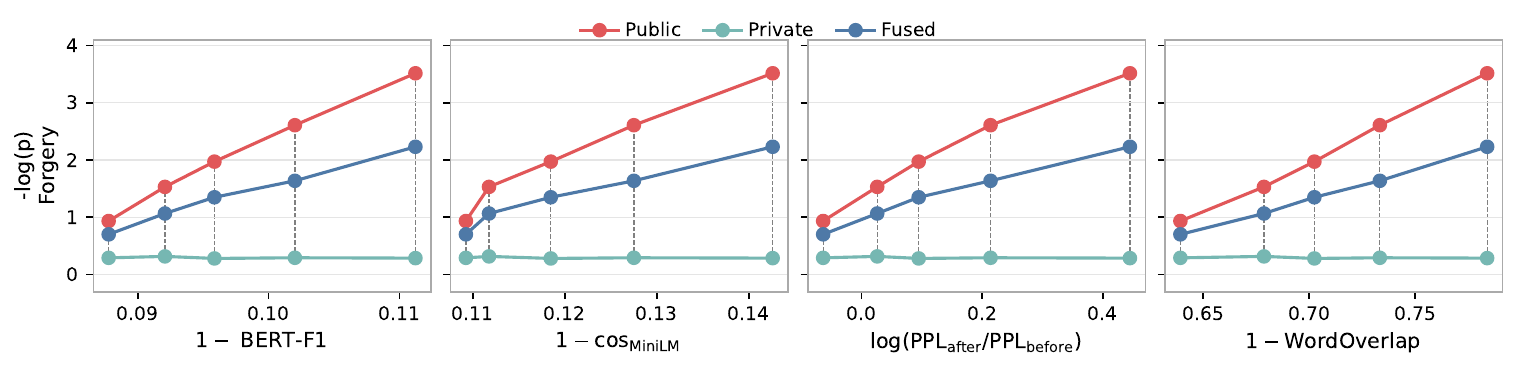}
\includegraphics[width=\linewidth]{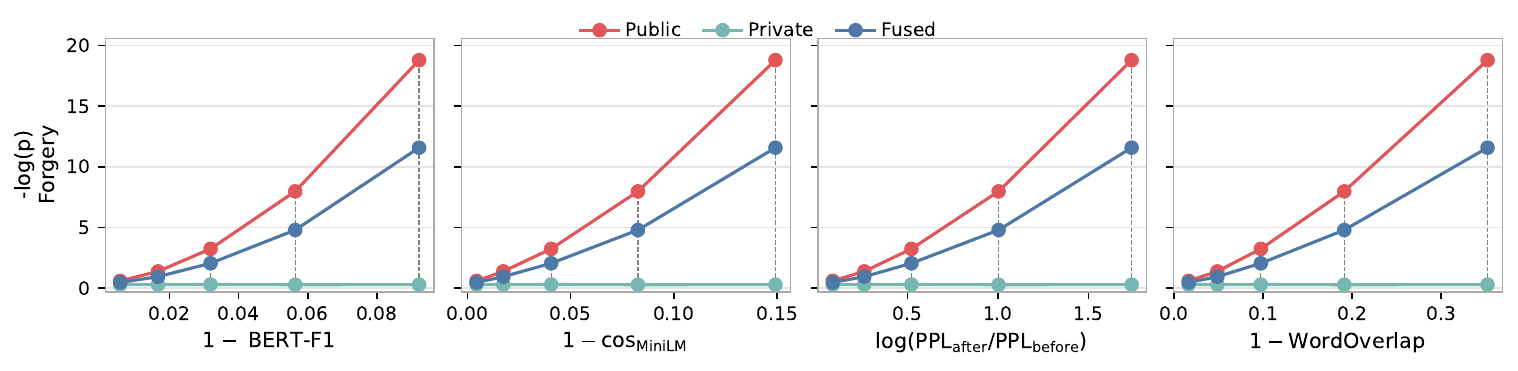}
\caption{
Forgery watermark score over quality degradation with rephrasing (top) and word edits (bottom).
}
\label{fig:pareto_more_quality_forgery}
\end{figure}

Regarding distortion quantification, we revisit the white-box attack sweeps of Fig.~\ref{fig:asr_distortion_channels} and measure their distortion using additional text quality metrics in Fig.~\ref{fig:pareto_more_quality_removal} for the removal goal, and Fig.~\ref{fig:pareto_more_quality_forgery} for the forgery goal.
From left to right, we present the distortion with: $1 - \text{BERT-F1}$ (the default metric used throughout), the cosine distance using MiniLM \citep{wang2020minilm}, the log perplexity ratio from GPT-2, and word overlap distance.

\begin{table}[t!]
\centering
\captionsetup{font=small}
\caption{
Per-attack removal results behind Fig.~\ref{fig:main_tamper}, over 1k attacked texts.
Success rates $\mathrm{ASR}$ are lower is better for the defender, and $\mathrm{ADR}$ and BERT-F1 are higher is better.
\colorbox{rowhl}{Light blue} marks the representative operating points reported in the main text (Fig.~\ref{fig:main_tamper}).
}
\label{tab:scrub_report}
\resizebox{\linewidth}{!}{
\begin{tabular}{lcccccccc}
\toprule
 & \multicolumn{3}{c}{Median $-\log p$ $\uparrow$} & Attacker & \multicolumn{3}{c}{Defender} & \multicolumn{1}{c}{Distortion} \\
\cmidrule(lr){2-4} \cmidrule(lr){5-5} \cmidrule(lr){6-8} \cmidrule(lr){9-9}
Attack & Public & Private & Fused & ASR\textsubscript{pub}$\downarrow$ & ASR\textsubscript{fus}$\downarrow$ & ADR$\uparrow$ & ASR\textsubscript{net}$\downarrow$ & $1-$BERT-F1$\downarrow$ \\
\midrule
Watermarked & 6.8 & 7.0 & 12.9 & 9.0\% & 1.7\% & 0.0\% & 1.7\% & 0.000 \\
\midrule
Rephrase (matched, no-box, add, T=0.5) & 1.5 & 1.7 & 2.5 & 79.1\% & 58.8\% & 0.2\% & 58.7\% & 0.039 \\
Rephrase (matched, no-box, add, T=0.8) & 1.2 & 1.3 & 1.8 & 89.2\% & 71.3\% & 0.3\% & 71.1\% & 0.045 \\
\rowcolor{rowhl} Rephrase (matched, no-box, add, T=1) & 0.9 & 1.0 & 1.5 & 92.3\% & 82.0\% & 0.4\% & 81.7\% & 0.052 \\
Rephrase (matched, no-box, add, T=1.2) & 0.8 & 0.8 & 1.1 & 97.5\% & 93.4\% & 0.1\% & 93.3\% & 0.060 \\
Rephrase (matched, no-box, add, T=1.5) & 0.5 & 0.6 & 0.7 & 99.2\% & 98.3\% & 0.4\% & 97.9\% & 0.074 \\
\rowcolor{rowhl} Rephrase (matched, white-box, add, T=0.5) & 0.4 & 1.3 & 1.0 & 98.2\% & 88.0\% & 1.9\% & 86.3\% & 0.046 \\
Rephrase (matched, white-box, add, T=0.8) & 0.0 & 0.9 & 0.2 & 100\% & 99.0\% & 12.9\% & 86.2\% & 0.057 \\
\rowcolor{rowhl} Rephrase (matched, white-box, add, T=1) & 0.0 & 0.7 & 0.0 & 100\% & 100\% & 36.6\% & 63.4\% & 0.066 \\
Rephrase (matched, white-box, add, T=1.2) & 0.0 & 0.6 & 0.0 & 100\% & 100\% & 70.3\% & 29.7\% & 0.079 \\
\rowcolor{rowhl} Rephrase (matched, white-box, add, T=1.5) & 0.0 & 0.4 & 0.0 & 100\% & 100\% & 93.0\% & 7.0\% & 0.103 \\
Rephrase (matched, white-box, Gumbel, T=0.5) & 0.0 & 1.2 & 0.4 & 99.4\% & 95.5\% & 13.7\% & 82.4\% & 0.063 \\
Rephrase (matched, white-box, Gumbel, T=0.8) & 0.0 & 0.9 & 0.0 & 100\% & 99.7\% & 57.3\% & 42.6\% & 0.073 \\
\rowcolor{rowhl} Rephrase (matched, white-box, Gumbel, T=1) & 0.0 & 0.7 & 0.0 & 100\% & 100\% & 87.8\% & 12.2\% & 0.081 \\
Rephrase (matched, white-box, Gumbel, T=1.2) & 0.0 & 0.6 & 0.0 & 100\% & 100\% & 97.8\% & 2.2\% & 0.093 \\
Rephrase (matched, white-box, Gumbel, T=1.5) & 0.0 & 0.5 & 0.0 & 100\% & 100\% & 99.8\% & 0.2\% & 0.145 \\
\midrule
Word edits (no-box, 2\%) & 6.1 & 6.3 & 11.6 & 12.4\% & 2.1\% & 0.0\% & 2.1\% & 0.004 \\
\rowcolor{rowhl} Word edits (no-box, 5\%) & 5.1 & 5.3 & 9.7 & 17.7\% & 3.3\% & 0.0\% & 3.3\% & 0.010 \\
Word edits (no-box, 10\%) & 4.0 & 4.1 & 7.4 & 31.7\% & 6.7\% & 0.0\% & 6.7\% & 0.019 \\
\rowcolor{rowhl} Word edits (no-box, 20\%) & 2.5 & 2.6 & 4.4 & 62.5\% & 26.9\% & 0.0\% & 26.9\% & 0.032 \\
Word edits (no-box, 40\%) & 1.2 & 1.2 & 1.8 & 91.8\% & 77.0\% & 0.0\% & 77.0\% & 0.055 \\
Word edits (white-box, 2\%) & 3.6 & 6.4 & 9.1 & 36.6\% & 4.7\% & 4.3\% & 4.5\% & 0.005 \\
\rowcolor{rowhl} Word edits (white-box, 5\%) & 1.2 & 5.5 & 5.2 & 91.5\% & 20.9\% & 12.9\% & 18.2\% & 0.013 \\
\rowcolor{rowhl} Word edits (white-box, 10\%) & 0.1 & 4.1 & 1.7 & 99.8\% & 79.3\% & 59.1\% & 32.4\% & 0.025 \\
\rowcolor{rowhl} Word edits (white-box, 20\%) & 0.0 & 2.4 & 0.1 & 99.9\% & 99.2\% & 97.1\% & 2.9\% & 0.046 \\
Word edits (white-box, 40\%) & 0.0 & 1.2 & 0.0 & 99.9\% & 99.2\% & 99.4\% & 0.6\% & 0.080 \\
\midrule
Average (25 attacks) & 0.2 & 1.4 & 0.7 & 83.9\% & 72.2\% & 33.8\% & 39.6\% & 0.053 \\
\bottomrule
\end{tabular}

}
\end{table}

\begin{table}[t!]
\centering
\captionsetup{font=small}
\caption{
Per-attack forgery results behind Fig.~\ref{fig:main_tamper}, over 1k attacked texts.
Columns follow Tab.~\ref{tab:scrub_report}, and \colorbox{rowhl}{light blue} again marks the representative operating points from the main text (Fig.~\ref{fig:main_tamper}).
}
\label{tab:spoof_report}
\resizebox{\linewidth}{!}{
\begin{tabular}{lcccccccc}
\toprule
 & \multicolumn{3}{c}{Median $-\log p$} & Attacker & \multicolumn{3}{c}{Defender} & \multicolumn{1}{c}{Distortion} \\
\cmidrule(lr){2-4} \cmidrule(lr){5-5} \cmidrule(lr){6-8} \cmidrule(lr){9-9}
Attack & Public & Private & Fused & ASR\textsubscript{pub}$\downarrow$ & ASR\textsubscript{fus}$\downarrow$ & ADR$\uparrow$ & ASR\textsubscript{net}$\downarrow$ & $1-$BERT-F1$\downarrow$ \\
\midrule
Unwatermarked & 0.3 & 0.3 & 0.3 & 0.1\% & 0.0\% & 0.0\% & 0.0\% & 0.000 \\
\midrule
\rowcolor{rowhl} Rephrase (matched, white-box, add, T=0.5) & 0.9 & 0.3 & 0.7 & 3.2\% & 1.0\% & 0.0\% & 1.0\% & 0.088 \\
Rephrase (matched, white-box, add, T=0.8) & 1.5 & 0.3 & 1.1 & 9.5\% & 3.8\% & 0.0\% & 3.8\% & 0.092 \\
\rowcolor{rowhl} Rephrase (matched, white-box, add, T=1) & 2.0 & 0.3 & 1.3 & 20.8\% & 7.5\% & 10.7\% & 6.7\% & 0.096 \\
Rephrase (matched, white-box, add, T=1.2) & 2.6 & 0.3 & 1.6 & 38.2\% & 12.6\% & 15.9\% & 10.6\% & 0.102 \\
\rowcolor{rowhl} Rephrase (matched, white-box, add, T=1.5) & 3.5 & 0.3 & 2.2 & 62.0\% & 29.5\% & 37.6\% & 18.4\% & 0.111 \\
Rephrase (matched, white-box, Gumbel, T=0.5) & 1.9 & 0.3 & 1.2 & 21.5\% & 8.1\% & 11.1\% & 7.2\% & 0.087 \\
Rephrase (matched, white-box, Gumbel, T=0.8) & 4.2 & 0.3 & 2.6 & 74.0\% & 40.2\% & 42.5\% & 23.1\% & 0.090 \\
\rowcolor{rowhl} Rephrase (matched, white-box, Gumbel, T=1) & 7.1 & 0.3 & 4.3 & 95.4\% & 75.4\% & 72.9\% & 20.4\% & 0.093 \\
Rephrase (matched, white-box, Gumbel, T=1.2) & 11.5 & 0.3 & 7.2 & 99.9\% & 94.2\% & 91.5\% & 8.0\% & 0.097 \\
Rephrase (matched, white-box, Gumbel, T=1.5) & 24.1 & 0.3 & 15.4 & 99.9\% & 99.7\% & 99.7\% & 0.3\% & 0.109 \\
\midrule
Word edits (white-box, 2\%) & 0.6 & 0.3 & 0.5 & 0.5\% & 0.5\% & 0.0\% & 0.5\% & 0.006 \\
\rowcolor{rowhl} Word edits (white-box, 5\%) & 1.4 & 0.3 & 0.9 & 7.8\% & 3.2\% & 6.2\% & 3.0\% & 0.017 \\
\rowcolor{rowhl} Word edits (white-box, 10\%) & 3.2 & 0.3 & 2.0 & 56.3\% & 22.8\% & 19.3\% & 18.4\% & 0.032 \\
\rowcolor{rowhl} Word edits (white-box, 20\%) & 8.0 & 0.3 & 4.8 & 99.3\% & 84.9\% & 77.9\% & 18.8\% & 0.056 \\
Word edits (white-box, 40\%) & 18.8 & 0.3 & 11.6 & 99.9\% & 99.7\% & 100\% & 0.0\% & 0.092 \\
\midrule
Average (15 attacks) & 3.2 & 0.3 & 2.1 & 52.5\% & 38.9\% & 39.0\% & 9.3\% & 0.088 \\
\bottomrule
\end{tabular}

}
\end{table}

\subsection{Generalization to Other Models}
\label{app:othermodels}

We repeat the main experiments with Gemma-3-4B-IT~\citep{team2025gemma3} as the text generator, on the same dataset and settings, with Gemma-3-1B-IT as the \textit{matched} proxy rephrasing model.
How much watermark signal a text carries depends on the generator's output entropy, and Gemma-3-4B is less entropic than Qwen2.5-7B, so at a shared temperature its texts start with far less signal to remove and no attack would be comparable. 
We therefore raise its sampling temperature to $1.5$ to share approximately the same $p$-value per 100 tokens as on Qwen, and attack from there.

Tab.~\ref{tab:gemma} shows that, once the two generators start from the same signal, the defense behaves the same way on a different model family: word edits and forgery reach the same rates as on Qwen, informed and uninformed alike. 
Rephrasing is the only exception because of the rephrasing model which is Gemma-1B in the case of Gemma-4B.
It rewrites much more heavily than the 3B proxy used on Qwen, and an attack that rewrites that much stops steering and starts behaving like an uninformed paraphrase (same mechanism as App.~\ref{app:crosstok}).
The table also includes SmolLM2-1.7B-Instruct, a smaller generator with bigger entropy, whose watermark is stronger at the same number of tokens.
For this model, we use SmolLM2-360M-Instruct for the rephrasing attack. 
Its watermark is the strongest of the three, so its absolute removal rates are the lowest, but the ordering between attacks stays unchanged.

\begin{table}[tb]
\centering
\captionsetup{font=small}
\caption{
Other generators compared to Qwen2.5-7B, at $p<10^{-3}$ on 1k texts. 
Gemma-3-4B generations have approximately the same watermark signal as Qwen per 100 tokens, so its rows are directly comparable. SmolLM2-1.7B is not, and carries a stronger watermark. 
The two-stage defense behaves the same way on all three.
}
\label{tab:gemma}
{\small
\begin{tabular}{llcccc}
\toprule
Attack & Generator & $\mathrm{ASR}_{\mathrm{fus}}$ & ADR & $\mathrm{ASR}_{\mathrm{net}}$ & Distortion \\
\midrule
\multirow{3}{*}{Rephrase, no-box, T${=}1$}      & Qwen2.5-7B   & 82.0\% & 0.4\%  & 81.7\% & 0.052 \\
                                                & Gemma-3-4B   & 93.7\% & 0.2\%  & 93.5\% & 0.107 \\
                                                & SmolLM2-1.7B & 91.7\% & 0.0\%  & 91.7\% & 0.097 \\
\midrule
\multirow{3}{*}{Rephrase, white-box, T${=}1$}   & Qwen2.5-7B   & 100\%  & 36.6\% & 63.4\% & 0.066 \\
                                                & Gemma-3-4B   & 98.8\% & 12.2\% & 86.7\% & 0.115 \\
                                                & SmolLM2-1.7B & 100\%  & 50.3\% & 49.7\% & 0.116 \\
\midrule
\multirow{3}{*}{Word edits, no-box, 10\%}       & Qwen2.5-7B   & 6.7\%  & 0.0\%  & 6.7\%  & 0.019 \\
                                                & Gemma-3-4B   & 8.0\%  & 0.0\%  & 8.0\%  & 0.019 \\
                                                & SmolLM2-1.7B & 3.4\%  & 0.0\%  & 3.4\%  & 0.017 \\
\midrule
\multirow{3}{*}{Word edits, white-box, 10\%}    & Qwen2.5-7B   & 79.3\% & 59.1\% & 32.4\% & 0.025 \\
                                                & Gemma-3-4B   & 68.6\% & 53.1\% & 32.2\% & 0.027 \\
                                                & SmolLM2-1.7B & 23.7\% & 26.2\% & 17.5\% & 0.023 \\
\midrule
\multirow{3}{*}{Forgery, edits, white-box, 10\%}& Qwen2.5-7B   & 22.8\% & 19.3\% & 18.4\% & 0.032 \\
                                                & Gemma-3-4B   & 20.1\% & 20.4\% & 16.0\% & 0.032 \\
                                                & SmolLM2-1.7B & 22.9\% & 14.0\% & 19.7\% & 0.031 \\
\bottomrule
\end{tabular}
}
\end{table}

\subsection{Watermark Strength Before Attack}
\label{app:prefix_pvalues}

Tab.~\ref{tab:prefix_pvalues} reports the three $p$-values the detector exposes: public, private and fused, before any attack, on the first $100$ tokens of each generation.
As expected, the two channels are interchangeable, while the fused $p$-value is lower.
The scores are modest because they are measured on a $100$-token prefix, whereas a full generation averages $360$ tokens and the fused test then flags $98.3\%$ of untampered watermarked texts at the same threshold (Tab.~\ref{tab:scrub_report}).
Across generators the watermark is much stronger on SmolLM2 than on Gemma~3 at equal length, as it directly depends on the entropy of the model (App.~\ref{app:othermodels}).
Raising Gemma's sampling temperature recovers most of the gap, which is why we generate with it at $T{=}1.5$ throughout.

\begin{table}[t!]
\centering
\captionsetup{font=small}
\caption{
Watermark strength before any attack, at $p<10^{-3}$ over 1k texts.
Both the median $-\log_{10} p$ and the TPR are measured on the first $100$ tokens of each generation, so that generators of different verbosity are compared at equal length.
Texts shorter than $100$ tokens are excluded from those columns.
$T$ is the sampling temperature used to generate, and ``Answer tokens'' the length of the full generation (mean $\pm$ std) in the generator's answers.
}
\label{tab:prefix_pvalues}
\small
\begin{tabular}{lcccccccc}
\toprule
 & & & \multicolumn{6}{c}{On the first $100$ tokens} \\
\cmidrule(lr){4-9}
 & & & \multicolumn{3}{c}{Median $-\log_{10} p$} & \multicolumn{3}{c}{TPR} \\
\cmidrule(lr){4-6} \cmidrule(lr){7-9}
Generator & $T$ & Answer tokens & $p_\mathrm{pub}$ & $p_\mathrm{priv}$ & $p_\mathrm{fus}$ & $p_\mathrm{pub}$ & $p_\mathrm{priv}$ & $p_\mathrm{fus}$ \\
\midrule
Qwen2.5-7B   & $1.0$  & $364 \pm 61$ & 2.0 & 2.0 & 3.3 & 24.6\% & 24.5\% & 57.7\% \\
Gemma-3-4B   & $1.5$ & $398 \pm 20$ & 1.4 & 1.5 & 2.2 & 10.3\% & 12.3\% & 30.2\% \\
Gemma-3-4B   & $1.0$  & $398 \pm 22$ & 0.9 & 0.9 & 1.3 & 1.6\%  & 2.8\%  & 6.4\%  \\
SmolLM2-1.7B & $1.0$  & $268 \pm 97$ & 4.6 & 4.5 & 8.3 & 75.7\% & 75.4\% & 96.3\% \\
\bottomrule
\end{tabular}
\end{table}

\subsection{Cross-Tokenizer Rephrasing}
\label{app:crosstok}

The rephrasing attacks so far use a proxy (the rephrasing model) sharing the generator's tokenizer. 
An attacker has no particular reason to own one: the tokenizer belongs to a model it does not have, and the obvious choice is whatever open model it already runs.
We call the attack \emph{matched} when the two tokenizers agree and \emph{cross} when they do not, and run two cross proxies against the Qwen2.5-7B generator.
At each decoding step, for every candidate $v$ in its nucleus, the attacker decodes $v$ to a string, appends it to what it has written so far, and submits the whole string. The detector re-tokenizes it in the generator's vocabulary and returns $r$ for the last generator token, under the $h$ generator tokens before it. The attacker then biases the logit of $v$ by $\log r$, exactly as in the matched case.
Tab.~\ref{tab:crosstok} shows that the token overlap decides the outcome. 
A proxy that tokenizes similarly to the generator can steer the score more effectively, and the tampering test then flags it.
A proxy that tokenizes differently cannot steer the score, so the attack reduces to an uninformed paraphrase in the extreme case of little overlap, the one case the tampering test cannot detect, and it thus reaches the highest net removal rate.

\begin{table}[tb]
\centering
\captionsetup{font=small}
\caption{Matched against cross-tokenizer rephrasing on the Qwen2.5-7B generator, removal at T${=}1$ on 1k texts. ``Token overlap'' is the fraction of the generator's token spans that the proxy reproduces, measured on the generator's own generations. Both the attack steering and the token score-based tampering detection are applicable only at this overlap.}
\label{tab:crosstok}
{\small
\begin{tabular}{llccccc}
\toprule
Attacker & Proxy & Token overlap & $\mathrm{ASR}_{\mathrm{fus}}$ & ADR & $\mathrm{ASR}_{\mathrm{net}}$ & Distortion \\
\midrule
\multirow{3}{*}{Uninformed} & Matched (Qwen2.5-3B) & 100\%  & 82.0\% & 0.4\% & 81.7\% & 0.052 \\
                            & Cross (SmolLM3-3B)   & 99.3\% & 98.5\% & 0.0\% & 98.5\% & 0.137 \\
                            & Cross (Gemma-3-4B)   & 90.3\% & 99.1\% & 0.0\% & 99.1\% & 0.079 \\
\midrule
\multirow{3}{*}{Informed}   & Matched (Qwen2.5-3B) & 100\%  & 100\%  & 36.6\% & 63.4\% & 0.066 \\
                            & Cross (SmolLM3-3B)   & 99.3\% & 99.9\% & 65.8\% & 34.2\% & 0.143 \\
                            & Cross (Gemma-3-4B)   & 90.3\% & 100\%  & 4.9\%  & \textbf{95.1\%} & 0.081 \\
\bottomrule
\end{tabular}
}
\end{table}

\subsection{Adaptive Black-Box Attacks}
\label{app:blackbox}

We present the adaptive black-box attacks, as done for the white-box attacks in Sec.~\ref{sec:tamper_aware}.
A black-box detector only answers ``watermarked or not'', so the attacker cannot use it to choose which words to edit or what to replace them with.
They make both choices blindly, as in the no-box attack, and use the verdict only to decide when to stop.
As a reminder, the adaptive attacker edits until the verdict flips, and with an overshoot factor $m$ it keeps editing until it has made $m$ times the number of edits needed for the flip, capped by the edit budget (full details in App.~\ref{app:attacks}).
Tab.~\ref{tab:blackbox_report} sweeps $m$ for each edit budget, with the same metrics as Tab.~\ref{tab:adaptive_report} on the same 1k texts.

First, overshooting past the verdict flip only helps when the budget is large enough: \eg at $5\%$ the three stopping rules give equivalent results, because the attack reaches the budget before the verdict flips.
Second, at every budget the most harmful attack is no-box, which spends the full budget, but it is also the one that introduces the most distortion.
However, at matched distortion a partial overshoot can beat uninformed attacks.
Adaptive black-box attacks therefore improve the ASR-distortion tradeoff over the non-adaptive ones, but only slightly.
$\mathrm{ADR}$ stays at $0.0\%$: none of these attacks touches one channel more than the other, so the tampering test never flags them, and the only useful defense is the fact that the test on the fused score is more powerful.

\begin{table}[tb]
\centering
\captionsetup{font=small}
\caption{
Black-box removal via word edits, at $p<10^{-3}$ over 1k texts, grouped by edit budget.
Within a group, $m$ is the overshoot factor: after the verdict flips at $e$ edits the attacker continues to $m {\cdot} e$, capped by the budget, so $m{=}1$ stops on the crossing and no-box spends the budget without using the verdict.
Bold marks the highest $\mathrm{ASR}_{\mathrm{net}}$.
$\mathrm{ADR}$ is $0.0\%$ everywhere: no stopping rule leaves the channel imbalance the tampering test looks for, so the fused score alone is what limits the attacker.
}
\label{tab:blackbox_report}
{\colorlet{heathue}{heatblue!25!white}
\small
\resizebox{\linewidth}{!}{

\begin{tabular}{l*{13}{C{0.75cm}}}
\toprule
Edit budget & \multicolumn{3}{c}{5\%} & \multicolumn{3}{c}{10\%} & \multicolumn{3}{c}{20\%} & \multicolumn{4}{c}{40\%} \\
\cmidrule(lr){2-4} \cmidrule(lr){5-7} \cmidrule(lr){8-10} \cmidrule(lr){11-14}
Metric & $m{=}1$ & $m{=}2$ & No-box & $m{=}1$ & $m{=}2$ & No-box & $m{=}1$ & $m{=}2$ & No-box & $m{=}1$ & $m{=}2$ & $m{=}3$ & No-box \\
\midrule
ASR\textsubscript{pub}$\downarrow$ & \hc{19.2} & \hc{16.9} & \hc{17.7} & \hc{35.7} & \hc{31.5} & \hc{31.7} & \hc{72.6} & \hc{64.5} & \hc{62.5} & \hc{99.1} & \hc{93.4} & \hc{94.9} & \hc{91.8} \\
ASR\textsubscript{fus}$\downarrow$ & \hc{1.8} & \hc{1.8} & \hc{3.3} & \hc{2.3} & \hc{3.0} & \hc{6.7} & \hc{4.6} & \hc{16.3} & \hc{26.9} & \hc{10.4} & \hc{54.5} & \hc{69.3} & \hc{77.0} \\
ADR$\uparrow$ & \cellcolor{heathue!100.0}0.0\% & \cellcolor{heathue!100.0}0.0\% & \cellcolor{heathue!100.0}0.0\% & \cellcolor{heathue!100.0}0.0\% & \cellcolor{heathue!100.0}0.0\% & \cellcolor{heathue!100.0}0.0\% & \cellcolor{heathue!100.0}0.0\% & \cellcolor{heathue!100.0}0.0\% & \cellcolor{heathue!100.0}0.0\% & \cellcolor{heathue!100.0}0.0\% & \cellcolor{heathue!100.0}0.0\% & \cellcolor{heathue!100.0}0.0\% & \cellcolor{heathue!100.0}0.0\% \\
ASR\textsubscript{net}$\downarrow$ & \hc{1.7} & \hc{1.7} & \hc{3.3} & \hc{2.3} & \hc{3.0} & \hc{6.5} & \hc{4.3} & \hc{15.6} & \hc{25.6} & \hc{10.3} & \hc{53.7} & \hc{68.4} & \hc{75.3} \\
Distortion$\downarrow$ & 0.008 & 0.008 & 0.010 & 0.015 & 0.016 & 0.019 & 0.022 & 0.029 & 0.032 & 0.022 & 0.042 & 0.052 & 0.055 \\
\bottomrule
\end{tabular}

}
}
\end{table}

\subsection{Distribution of Flipping Distortion}
\label{app:flip_dist}

\begin{figure}[b!]
\centering
\includegraphics[width=0.5\linewidth]{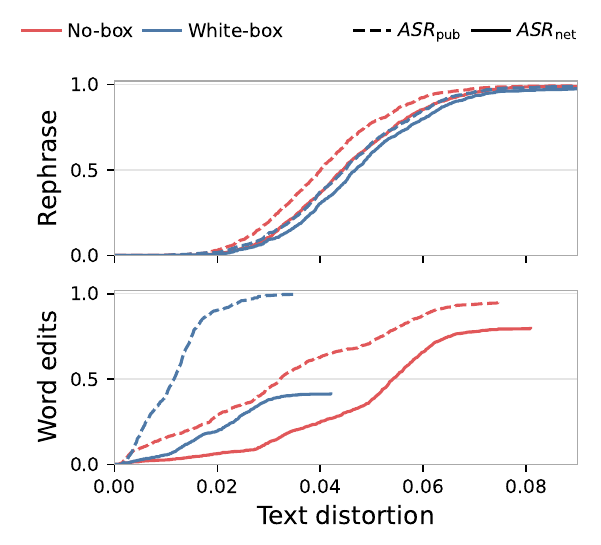}
\caption{%
Same sweep as the left panel of Fig.~\ref{fig:asr_distortion}, viewed per text: empirical CDF of the smallest flipping distortion for removal, no-box vs.\ white-box, rephrasing (top) and word edits (bottom).
White-box rephrasing barely improves over no-box at equal distortion, while white-box word edits flip the public detector at low distortion but rarely survive the two-stage test.
}
\label{fig:app_flip_cdf}
\end{figure}

As an alternative to the per-knob averages of Fig.~\ref{fig:asr_distortion} (left), we plot the empirical distribution of flipping distortion, following \citet{imadache2025evaluating}.
For each text $i$ we sweep the attack strength ($T \in \{0.5, 0.8, 1.0, 1.2, 1.5\}$ for rephrasing and edit fractions $\{2\%, 5\%, 10\%, 20\%, 40\%\}$ for word edits), and measure the text distortion $d$ as in the main text.
Let $d_i^c$ be the smallest distortion over the sweep at which decision $c$ flips ($d_i^c = \infty$ if it never flips), for $c \in \{\mathrm{pub}, \mathrm{net}\}$ where $\mathrm{net}$ additionally requires passing the tampering test.
We plot $\mathrm{CDF}_c(d) = \frac{1}{N} \sum_{i=1}^N\mathbbm{1}[d_i^c \leq d]$, the share of all $N$ texts flipped at distortion budget $d$, with texts that never flip keeping the curve below $1$.
Results for removal are in Fig.~\ref{fig:app_flip_cdf}, which gives the distortion needed to reach a given success rate.
Informed rephrasing and the uninformed baseline curves nearly coincide, so detector access gives the attacker little advantage there.
Informed word edits instead reach a public ASR near $100\%$ at distortions where rephrasing achieves almost nothing, but the two-stage test limits their net rate near $40\%$.

\subsection{A Gray-box Public Detector}
\label{app:less_transprent}

We now assume a gray-box detector that only returns the public-key $p$-value of the text.
We can directly adapt our white-box attack suite to read the score of a token from two detector queries.
The returned $p$-value is a decreasing function of the accumulated score $\mathbf{S}$ of the text, so the attacker inverts the watermark's null to recover $\mathbf{S}$, and takes $\mathbf{S}_{\text{w/}\ v} - \mathbf{S}_{\text{w/o}\ v}$ as the score of token $v$.
We examine the difference in the score space, not the $p$-values, because $p$ depends on the number $n$ of scored positions, which differs between the two texts.
This estimate is noisy for word edits, since removing $v$ also re-seeds the $h$ positions that follow it: their tokens are unchanged but their contexts are not, so their scores are redrawn and the difference carries that change along with the score of $v$.
On watermarked text the redrawn positions lose the watermark, but for small $h$ over large $n$ the effect on the total scores is small.
This noise even vanishes for rephrasing, where candidates are appended and the preceding context is fixed.

However, exploiting the gray-box detector means submitting numerous queries to it (at least twice as many as the white-box), which can be practically defended with rate-limiting.
Rephrasing attacks need to select the current token based on its score, so at every generation step they can conservatively query the detector at the top $k$ probable tokens.
Avoiding full regeneration with word edits results in fewer queries, since only a few words are going to be replaced.
However, this is more expensive in the case of gray-box, where the attacker needs to probe each token's importance by removing it from the text and comparing the difference in overall $p$-value.

\begin{figure}[b!]
\centering
\centering
\includegraphics[width=0.8\linewidth]{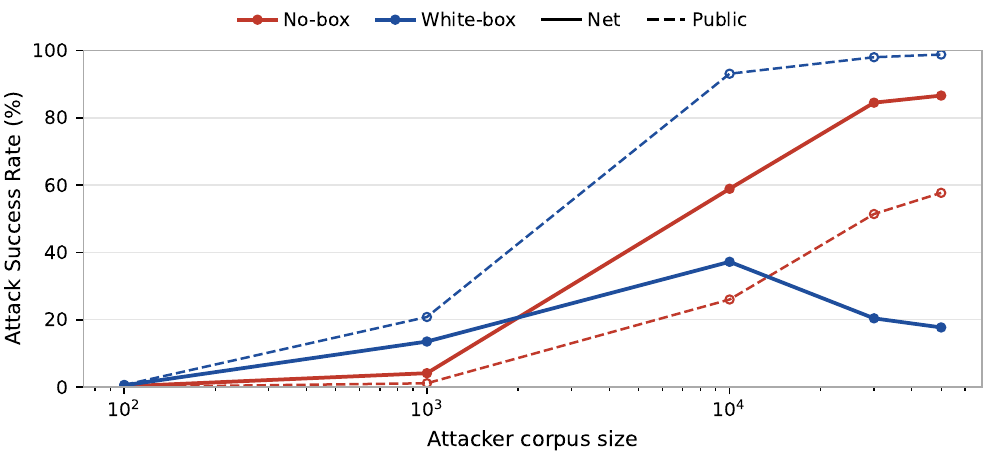}
\caption{Watermark stealing. Using the public detector increases the public ASR but not $\mathrm{ASR}_{\mathrm{net}}$, which stays no higher than the no-box attack, since supervision reaches only the public channel. The generation parameters are identical to previous experiments, with the exception of using $h=2$ as a context window, since our experiments with $h=3$ achieved at most 3\% ASR.}
\label{fig:learnability}
\hfill
\end{figure}

\subsection{Watermark Stealing}
\label{app:learnability}

So far, our experiments exploit the public detector in an online setting, querying it during tampering.
In this section, we examine watermark stealing methods \citep{jovanovic2024watermark, chen2025mark}, which learn an approximate watermark rule from a corpus, then perform offline attacks based on the reverse-engineered watermark.
We evaluate how advantageous the public detector can be for stealing attacks.
We use the purely statistical method of \citet{jovanovic2024watermark} as a baseline in the no-box scenario, which uses the frequency ratio of \texttt{(context, token)} pairs between a watermarked and unwatermarked corpus to estimate a per-token context-based bias.
We create a white-box adaptation, only changing the frequency estimation by using exactly the score for that pair instead of counting every appearance in the watermarked corpus equally.
The rephrasing attacks with the stolen watermark are performed with the same settings as the main experiments.

We present the stealing results in increasing sample budget in Fig.~\ref{fig:learnability}, showing that while detector exploitation accelerates the learning, it leads to very low $\mathrm{ASR}_{\mathrm{net}}$.
This happens because while no-box stealing extracts information from both the public and the private channels, exact supervision from the public detector only targets the former, causing channel score imbalance.
Using uninformed stealing without querying the detector is therefore more advantageous to an attacker who intends to reverse-engineer the watermark.
The stealing liability still exists, but is caused merely by the deployment of the watermark itself, not the public detector.

\subsection{Other Datasets}
\label{app:benchmarks}

We test whether the two-stage pipeline survives a change of language and script on CaLMQA \citep{arora2025calmqa}, a multilingual and culturally-specific question answering dataset.
We prompt Qwen2.5-7B with 1k questions from the dataset to generate watermarked responses towards removal, and unwatermarked responses towards forgery (since there are no human-written reference answers).
We modify our attacks by using a rephrasing prompt translated in the target language,
and substituting a multilingual version of RoBERTa \citep{conneau2020unsupervised} for word edits.
Because the 1k questions are split across languages, each row examines less than 100 documents, so per-language rates are expected to be noisier than the main experiments. 
Results are given in Tab.~\ref{tab:calmqa}.

\begin{table}[tb]
\centering
\captionsetup{font=small}
\caption{
CaLMQA by language at $p<10^{-3}$.
The median $-\log_{10} p$ columns are measured on unattacked watermarked text, and the attack columns report white-box word edits at $20\%$, the strongest arm of Sec.~\ref{subsec:attacks}.
$\mathrm{ASR}_{\mathrm{pub}}$ is what the attacker sees, $\mathrm{ADR}$ the percentage of tampering detections on texts that evade the fused detector, and $\mathrm{ASR}_{\mathrm{net}}$ what evades both stages.
We see that both the split watermark construction and the tampering test remain effective in low-resource languages.
}
\label{tab:calmqa}
\resizebox{\linewidth}{!}{
\begin{tabular}{llcccccccccccc}
\toprule
 & & & \multicolumn{3}{c}{Median $-\log_{10} p$ $\uparrow$} & \multicolumn{4}{c}{Removal} & \multicolumn{4}{c}{Forgery} \\
\cmidrule(lr){4-6}\cmidrule(lr){7-10}\cmidrule(lr){11-14}
Language & Script & $N$ & Public & Private & Fused & ASR\textsubscript{pub}$\downarrow$ & ASR\textsubscript{fus}$\downarrow$ & ADR$\uparrow$ & ASR\textsubscript{net}$\downarrow$ & ASR\textsubscript{pub}$\downarrow$ & ASR\textsubscript{fus}$\downarrow$ & ADR$\uparrow$ & ASR\textsubscript{net}$\downarrow$ \\
\midrule
\multicolumn{14}{l}{Higher-resource} \\
Hungarian & Latin & 72 & 17.5 & 18.0 & 34.2 & 98.6\% & 19.4\% & 42.9\% & 11.1\% & 86.1\% & 52.8\% & 47.4\% & 27.8\% \\
Chinese & Han & 72 & 10.2 & 10.7 & 20.5 & 100\% & 98.6\% & 94.4\% & 5.6\% & 100\% & 97.2\% & 90.0\% & 9.7\% \\
German & Latin & 72 & 9.9 & 10.7 & 19.2 & 100\% & 90.3\% & 84.6\% & 13.9\% & 94.4\% & 77.8\% & 69.6\% & 23.6\% \\
Arabic & Arabic & 72 & 9.4 & 10.2 & 18.3 & 100\% & 87.5\% & 77.8\% & 19.4\% & 91.7\% & 80.6\% & 77.6\% & 18.1\% \\
Hebrew & Hebrew & 65 & 8.5 & 9.5 & 17.1 & 100\% & 93.8\% & 68.9\% & 29.2\% & 95.4\% & 84.6\% & 85.5\% & 12.3\% \\
Korean & Hangul & 72 & 8.9 & 9.0 & 17.0 & 100\% & 87.5\% & 63.5\% & 31.9\% & 94.4\% & 75.0\% & 64.8\% & 26.4\% \\
Japanese & Japanese & 72 & 8.1 & 8.6 & 15.3 & 100\% & 100\% & 95.8\% & 4.2\% & 100\% & 97.2\% & 92.9\% & 6.9\% \\
English & Latin & 71 & 7.3 & 8.3 & 15.2 & 100\% & 100\% & 98.6\% & 1.4\% & 97.2\% & 95.8\% & 92.6\% & 7.0\% \\
Spanish & Latin & 72 & 7.2 & 6.7 & 13.1 & 100\% & 100\% & 90.3\% & 9.7\% & 100\% & 87.5\% & 90.5\% & 8.3\% \\
Russian & Cyrillic & 72 & 6.8 & 5.9 & 11.6 & 100\% & 97.2\% & 75.7\% & 23.6\% & 94.4\% & 65.3\% & 66.0\% & 22.2\% \\
Hindi & Devanagari & 72 & 3.1 & 3.8 & 6.1 & 100\% & 97.2\% & 34.3\% & 63.9\% & 37.5\% & 5.6\% & 25.0\% & 4.2\% \\
\midrule
\multicolumn{14}{l}{Lower-resource} \\
Hiligaynon & Latin & 48 & 22.0 & 21.8 & 42.4 & 100\% & 27.1\% & 61.5\% & 10.4\% & 97.9\% & 81.2\% & 74.4\% & 20.8\% \\
Kirundi & Latin & 39 & 15.1 & 17.1 & 35.1 & 59.0\% & 33.3\% & 61.5\% & 12.8\% & 71.8\% & 35.9\% & 42.9\% & 20.5\% \\
Tswana & Latin & 48 & 17.3 & 17.1 & 34.4 & 95.8\% & 43.8\% & 14.3\% & 37.5\% & 77.1\% & 60.4\% & 51.7\% & 29.2\% \\
Pashto & Arabic & 56 & 15.8 & 15.6 & 32.4 & 94.6\% & 32.1\% & 38.9\% & 19.6\% & 69.6\% & 50.0\% & 28.6\% & 35.7\% \\
Samoan & Latin & 18 & 8.0 & 8.4 & 16.6 & 100\% & 55.6\% & 40.0\% & 33.3\% & 77.8\% & 66.7\% & 58.3\% & 27.8\% \\
Papiamento & Latin & 7 & 4.4 & 5.5 & 7.3 & 100\% & 85.7\% & 83.3\% & 14.3\% & 71.4\% & 57.1\% & 50.0\% & 28.6\% \\
\midrule
Average & -- & -- & 9.1 & 9.4 & 17.5 & 97.8\% & 77.3\% & 74.4\% & 19.8\% & 87.6\% & 70.9\% & 74.9\% & 17.8\% \\
\bottomrule
\end{tabular}

}
\end{table}

\section{Asymmetric Contexts}
\label{app:asym}

So far both keys share one $h$-gram context.
The tampering test then uses a single signal, the public/private gap $\mathbf{T} = \mathbf{S}_{\mathrm{pub}} - \mathbf{S}_{\mathrm{priv}}$, which moves only when an attack treats the two keys differently.
A no-box paraphrase degrades both channels alike and leaves $\mathbf{T}$ near zero, which is why the uninformed attacker remains the stronger removal threat (Fig.~\ref{fig:main_tamper}).

\begin{wrapfigure}[17]{r}{0.43\textwidth}
\centering
\includegraphics[width=0.9\linewidth]{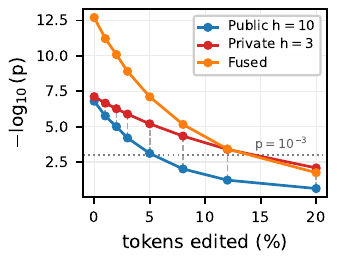}
\captionsetup{skip=4pt}
\caption{Idealised i.i.d.\ token edits at $(h_{\mathrm{pub}},h_{\mathrm{priv}}){=}(10,3)$: the wide public context decays faster, so the gap grows with editing. Not enough, though: net success stays worse than at equal contexts.}
\label{fig:asym_decay}
\end{wrapfigure}

Unequal contexts are a natural way to close that gap, however, we report them as a negative result.
A channel keeps its signal at a token only if the whole $(h{+}1)$-gram seeding its pseudorandom function survives the edit, with probability $q_h = (1-\rho)^{h+1}$ at edit rate $\rho$, so a wide context on the public key decays faster than a short one on the private key and uninformed editing does move $\mathbf{T}$ (Fig.~\ref{fig:asym_decay}).

The contrast becomes visible where equal contexts leave it at zero, and the sign-flip null of Sec.~\ref{subsec:tamper} calibrates it unchanged.
However, the wide public channel carries half the fused score and rephrasing destroys it, so the fused test gives up more detection than the tampering test recovers, at every amount of text we scored.
The asymmetry advantage is merely forensic, the ability to say that a corpus was rephrased.
Concurrent work by~\citet{feng2026tracing} similarly pairs a robust with a fragile signal, fixing its tamper threshold at the first percentile of the fragile score on intact watermarked text, where our sign-flip null needs no such corpus.

\begin{table}[t!]
\centering
\small
\caption{$\mathrm{ASR}_{\mathrm{net}}$ (\%, lower is better for the provider) against the number of tokens scored, with both tests run over all of it, at $p<10^{-3}$.
Bold marks the better configuration per column.
The tampering test does flag some of the asymmetric runs, with $\mathrm{ADR}$ rising to about $40\%$, but it recovers less than the fused test gives up.}
\label{tab:pool_both}
\begin{tabular}{rrrrrrr}
\toprule
 & \multicolumn{2}{c}{$T=0.5$} & \multicolumn{2}{c}{$T=0.8$} & \multicolumn{2}{c}{$T=1.2$} \\
\cmidrule(lr){2-3}\cmidrule(lr){4-5}\cmidrule(lr){6-7}
Scored tokens & $(3,3)$ & $(10,3)$ & $(3,3)$ & $(10,3)$ & $(3,3)$ & $(10,3)$ \\
\midrule
296 & \textbf{65.3} & 78.8 & \textbf{75.2} & 89.5 & \textbf{93.8} & 96.2 \\
592 & \textbf{31.7} & 55.4 & \textbf{51.5} & 74.6 & \textbf{83.4} & 93.2 \\
1.5k & \textbf{2.1} & 16.4 & \textbf{10.3} & 37.5 & \textbf{50.9} & 84.2 \\
3.0k & \textbf{0.0} & 0.8 & \textbf{0.2} & 7.0 & \textbf{14.8} & 56.5 \\
7.4k & \textbf{0.0} & 0.0 & \textbf{0.0} & 0.0 & \textbf{0.1} & 8.5 \\
14.8k & \textbf{0.0} & 0.0 & \textbf{0.0} & 0.0 & \textbf{0.0} & 0.0 \\
\bottomrule
\end{tabular}

\end{table}

\paragraph{Placement.}
The wide context goes on the public key.
An edit lowers the public score and also re-seeds the $h$ positions after it, so both effects hit the same channel.
Reversing the placement sends the second effect to the private channel, where it cancels part of the first.

\paragraph{Calibration.}
At detection, a position is kept only if its context is new in every channel, so that all channels are scored on the same positions.
On $5992$ untampered generations at $(10,2)$ the pooled $z$ is then $-0.001 \pm 0.013$, and the per-document FPR is $0.0017$, $0.011$ and $0.053$ for nominal $10^{-3}$, $0.01$ and $0.05$.

\paragraph{Uninformed attacker.}
We apply the no-box rephrasing of Sec.~\ref{subsec:attacks} to $(3,3)$ and to $(10,3)$, $1000$ documents per temperature, at matched quality (BERT-F1 $0.952$ at $T=0.8$ for both).
At $300$ scored tokens the asymmetric design is already worse: the public channel is wide, rephrasing destroys it, and it carries half the fused score.

The contrast is a sum, so $|z|$ grows with the square root of the text scored.
Over the removed text it reaches $-24.3$, against $-0.6$ for the control.
At $T=0.8$ it flags about $1\%$ at $300$ scored tokens, $29\%$ at $3\times10^{3}$ and $99\%$ at $1.6\times10^{4}$, while the control stays at its nominal rate.

That sensitivity does not lower the net rate.
Tab.~\ref{tab:pool_both} runs both tests over increasing amounts of scored text, showing that equal contexts are better at every size.
Rephrasing destroys the wide public channel, so the fused test loses more detection than the contrast recovers, and the asymmetry adds forensic evidence alone.

\end{document}